\documentclass{article}

\usepackage{PRIMEarxiv}

\usepackage[utf8]{inputenc} 
\usepackage[T1]{fontenc}    
\usepackage{hyperref}       
\usepackage{url}            
\usepackage{booktabs}       
\usepackage{amsfonts}       
\usepackage{nicefrac}       
\usepackage{microtype}      
\usepackage{lipsum}
\usepackage{fancyhdr}       
\usepackage{graphicx}       
\graphicspath{{media/}}     
\usepackage{amsmath}
\usepackage{amssymb}
\usepackage{mathrsfs}  
\usepackage[numbers]{natbib}
\title{Responses to Disturbance of Supersonic Shear Layer: Input-Output Analysis} 

\author{
  Mitesh Thakor\thanks{\texttt{corresponding author}}, Yiyang Sun \\
  Department of Mechanical and Aerospace Engineering\\
  Syracuse University\\
  Syracuse, NY 13210 \\
  \texttt{\{mmthakor, ysun58\}@syr.edu} \\
   \And
  Datta V. Gaitonde \\ 
  Department of Mechanical and Aerospace Engineering\\
  The Ohio State University\\
  Columbus, OH 43210\\
  \texttt{gaitonde.3@osu.edu} \\
}

\begin{document}
\maketitle

\begin{abstract}

We investigate the perturbation dynamics in a supersonic shear layer using a combination of large-eddy simulations (LES) and linear-operator-based input-output analysis. The flow consists of two streams -- a main stream (Mach~1.23) and a bypass stream (Mach~1.0) -- separated by a splitter plate of non-negligible thickness. We employ spectral proper orthogonal decomposition to identify the most energetic coherent structures and bispectral mode decomposition to explore the nonlinear energy cascade within the turbulent shear layer flow. Structures at the dominant frequency are also obtained from a resolvent analysis of the mean flow. We observe higher gain at the dominant frequency in resolvent analysis, indicating the dominance of Kelvin--Helmholtz (KH) instability as the primary disturbance energy-amplification mechanism. To focus on realizable actuator placement locations, we further conduct an input-output analysis by restricting a state variable and spatial location of an input and output. Various combinations of inputs and output indicate that the splitter plate trailing surface is the most sensitive location for introducing a perturbation. Upper and lower surface inputs are less influential in modulating wavepackets in the shear layer but introduce pressure instability waves in the main and bypass streams, respectively. The analysis reveals that the phase speed of pressure waves depends on the state variable and input location combination. For all combinations, the KH instability plays a key role in amplification, which reduces significantly as the input location is moved upstream relative to the splitter plate trailing edge. Furthermore, two-dimensional nonlinear simulations with unsteady input at the upper surface of the splitter plate show remarkable similarities between pressure modes obtained through dynamic mode decomposition and those predicted from linear input-output analysis at a given frequency. This study emphasizes the strength of linear analysis and demonstrates that predicted coherent structures remain active in highly nonlinear turbulent flow. The insights gained from the input-output analysis can be further leveraged to formulate practical flow control strategies.

\end{abstract}


\section{Introduction} \label{sec:intro}

\begin{figure}
     \centering
         \includegraphics[width=0.8\textwidth]{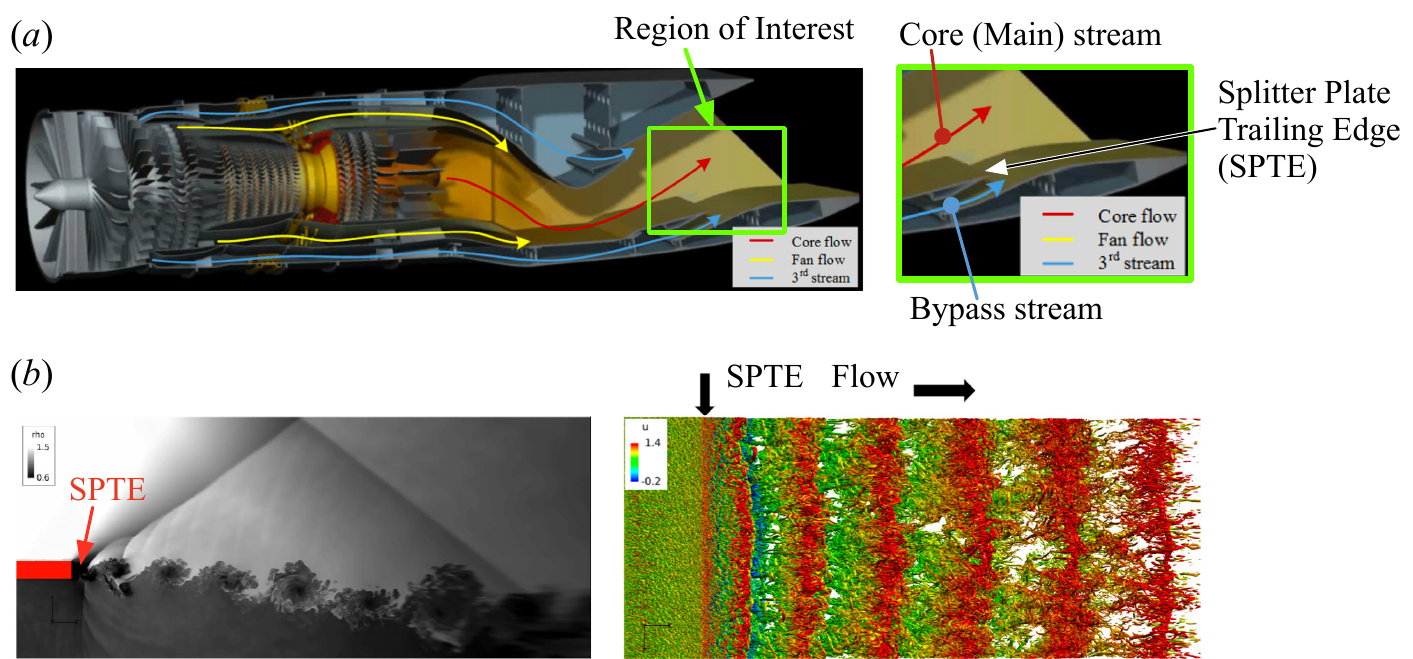}
    \caption{(a) Three-stream turbofan engine architecture developed by Air Force Research Laboratory (AFRL) discussed by \citet{simmons2009}. (b) Instantaneous density and the Q-criterion colored by the streamwise velocity component of the supersonic shear layer by \citet{stack2019splitter}.}
    \label{fig:engine}
\end{figure}

Jet engines have undergone significant evolution, progressing from relatively simple configurations to their advanced modern state. This evolution has been driven by the need for higher thrust generation, as well as the demand for providing sufficient energy to support advanced avionics and meet rigorous mission-based requirements. A three-stream non-axisymmetric, airframe-integrated, variable-cycle engine described by \citet{simmons2009} is shown in figure~\ref{fig:engine}(a). The design incorporates a rectangular single expansion ramp nozzle (SERN) consisting of a core and two bypass streams to fulfill performance requirements. This design allows better integration with a slender aircraft frame, hence reducing drag (\citet{capone1979}), and enhances propulsion efficiency under various flight conditions. A core and primary bypass (fan) stream produces power, while a secondary bypass (third) stream is utilized to provide a thermal cooling bed to an aircraft frame from a hot core stream (\citet{bruening1999}) and to reduce overall noise generation (\citet{papamoschou2001, magstadt2015, berry2016}).

The above-mentioned nozzle configuration has been extensively studied both numerically (\citet{stack2018, stack2019}) and experimentally (\citet{magstadt2015, magstadt2017, berry2017, berry2017low}). These studies have reported the formation of large coherent structures resulting from the mixing of the core (or ``main'') and secondary (``bypass'') streams, leading to a vortex-shedding instability at the trailing edge of the splitter plate. These large coherent structures are responsible for producing high surface loading on the aircraft frame and generating a strong noise signature in the far field. A simplified model of isolated shear-layer flow has been used to investigate the underlying mechanism of the flow, consisting only of the splitter plate and the main stream and bypass streams, at the same conditions as in the full configuration (\citet{stack2019splitter, doshi2022}) (see figure~\ref{fig:engine}(b)). This simplification successfully isolates the genesis of the main phenomena of interest for detailed study; including the formation and downstream evolution of coherent structures and their implications on the unsteady shock system. In this paper, we use this isolated shear-layer model to extend the analysis further to encompass the input-output dynamics of perturbations, which can provide insights for altering the undesired features of the nozzle flow for an improved design of the nozzle component of the jet engine.

Canonical supersonic shear layers or mixing layers have been extensively examined due to their fundamental relevance in supersonic combustion, high-speed civil transport, and other applications. The mean velocity profile of a shear layer exhibits an inflection point, leading to the Kelvin--Helmholtz instability and the formation of large-scale coherent structures. The spatiotemporal dynamics of these coherent structures have been extensively studied, including their growth rate, vortical structures, acoustic interactions, and shock waves, across various Mach numbers (\citet{bogdanoff1983, papamoschou1988, gutmark1995, delville1999, ukeiley2001}). Moreover, linear stability theory indicates that a convective Mach number (\citet{papamoschou1988}) greater than 0.6 amplifies three-dimensional disturbances (\citet{sandham1990}), a value significantly higher than our study (approximately 0.087). Therefore, we focus on the dynamics of two-dimensional disturbances in this study. Further, the utilization of small amplitude unsteady forcing to modulate shear layer behavior through its natural instability mechanism has been well explored (\citet{greenblatt2000}). Flow instabilities facilitate the growth of small-scale disturbances by an order of magnitude or more through their natural amplification mechanism, which can be harnessed to develop a low-amplitude yet effective scalable actuator.

Numerous efforts have been made to control shear-layer instabilities and their subsequent growth through passive flow control, primarily by modifying the geometry of the splitter plate (\citet{gutmark1995}). The shear layer of interest in this work, figure~\ref{fig:engine}, is distinct in that the properties of the two independent streams on either side of the splitter plate are influenced by flow events inside the engine and, as such, are not strictly in equilibrium prior to mixing. The splitter plate is also relatively thick, of an order of magnitude greater than the boundary layer height, and thus is susceptible to shedding instability as well. \citet{stack2019splitter} investigated the influence of splitter plate thickness on the length scale of coherent structures and instability growth. They found that as the thickness of the plate decreases, the instability mechanism changes.  At one-tenth the actual thickness,  the coherent structures were greatly minimized and the impact of the splitter plate on the rest of the flow field was relatively small. Similarly, \citet{ruscher2018} observed a transition from vortex-shedding-like to Kelvin-Helmholtz-like instability with a knife-edged plate instead of a flat splitter plate. Guided by the linear analysis, \citet{doshi2022} conducted high-fidelity simulations with a wavy splitter plate trailing edge for an isolated shear layer. They demonstrated a breakdown of spanwise two-dimensional coherent roller vortical structures into small-scale structures and their energy redistribution through data-driven techniques. Experiments with the same wavy splitter plate for a full nozzle configuration confirmed a reduction in the dominant peak of far-field acoustic spectra (\citet{gist2020}). 

While passive flow control designs have demonstrated promising outcomes, they might degrade for off-design operating conditions. On the other hand, active flow control is more flexible and may be tailored for different operating conditions. Nevertheless, developing an optimal design is complicated due to the prohibitive parameter space, which includes actuator placement, geometry, waveform characteristics, and amplitudes. To circumvent the tedious trial-and-error process of active flow design through experimentation or high-fidelity numerical simulation, we employ an operator-based input-output analysis (or classical resolvent analysis), augmented with select numerical studies, to gain physical insight into perturbation behaviors that may be leveraged into a coherent strategy. 

In the past decade, linear-operator-based resolvent analysis has emerged as a highly effective tool for extracting crucial features from fluid flows (\citet{farrell1993}). This method is particularly appealing for flow control designs as it identifies the optimal input capable of amplifying the flow response effectively at a specific frequency (\citet{trefethen1993, mckeon2010}). The classical resolvent analysis has been successfully applied to various flow configurations, including open cavity flows (\citet{gomez2016, sun2017, liu2018}), jet flows (\citet{schmidt2018, semeraro2016Stochastic, lesshafft2019, towne2017}), airfoil flows (\citet{symon2019, yeh2019}), flow over backward-facing steps (\citet{dergham2013}), and channel flows (\citet{moarref2014}). By considering all state variables, resolvent analysis reveals the overall amplification mechanisms of the flow. However, in flow control applications, actuators are typically located near solid surfaces and introduce inputs in specific forms of forcing (\citet{cattafesta2011}), such as momentum- (\citet{glezer2002}), acoustic- (\citet{heidt2021}), or thermal-based (\citet{little2019}) disturbances. To model such control-oriented configurations, the classical resolvent analysis has been extended to examine componentwise amplification mechanics for particular input-output scenarios, with spatial or variable restrictions (\citet{jovanovic2005, jeun2016, yeh2019}). In this study, we isolate and analyze the supersonic shear-layer flow associated with the splitter plate of the multistream engine. The global perturbation dynamics is examined using the classical resolvent analysis considering full-state variables without spatial restriction, and an input-output analysis is used with state variables and spatial restrictions to gain insights that lead to more targeted efforts in active flow control strategies.

The paper is structured as follows: The physical model problem of the supersonic shear layer is presented in \S~\ref{sec:phy_model}, followed by an explanation of the framework for input-output analysis in \S~\ref{sec:inout}. This section covers both the classical resolvent and input-output configurations, along with a comparison of full singular value decomposition (SVD) and randomized SVD. In \S~\ref{sec:result}, we first characterize the baseline shear layer flow and then employ the data-driven modal analysis technique, namely spectral proper orthogonal decomposition (SPOD), to gain further insight into spatiotemporal dynamics. Subsequently, the results of classical resolvent analysis are presented and compared with SPOD. Additionally, we elucidate the amplification mechanism and modal structures for various combinations of the input variable and spatial location. The validity of input-output analysis is tested against nonlinear simulations with unsteady forcing by comparing the output modes to dynamic mode decomposition results for two forcing cases in \S~\ref{sec:2D_simulaiton}. Finally, detailed conclusions and remarks on the present study are provided in \S~\ref{sec:conclusion}.

\section{Methods}

\subsection{Physical Model Problem} \label{sec:phy_model}

In the present work, we consider a simplified version of the complex multi-stream nozzle flow of figure~\ref{fig:engine} in the region of the splitter plate. Of specific interest is the shear layer forming downstream of the splitter plate that separates the core and bypass streams. A three-dimensional (3-D) unsteady large-eddy simulation (LES) was performed for the shear-layer flow over a thick splitter plate by \citet{stack2019splitter}.  The results of figure~\ref{fig:engine}(b) highlight the growth of coherent structures downstream of the splitter plate with their general downward inclination due to the flow parameters of the core and bypass streams (in table~\ref{tab:1}).  In the initial region downstream of the plate, the top view (right) indicates the relatively two-dimensional evolution of the structures with increasing spanwise breakdown appearing relatively far downstream.

\begin{figure}
     \centering
         \includegraphics[width=0.85\textwidth]{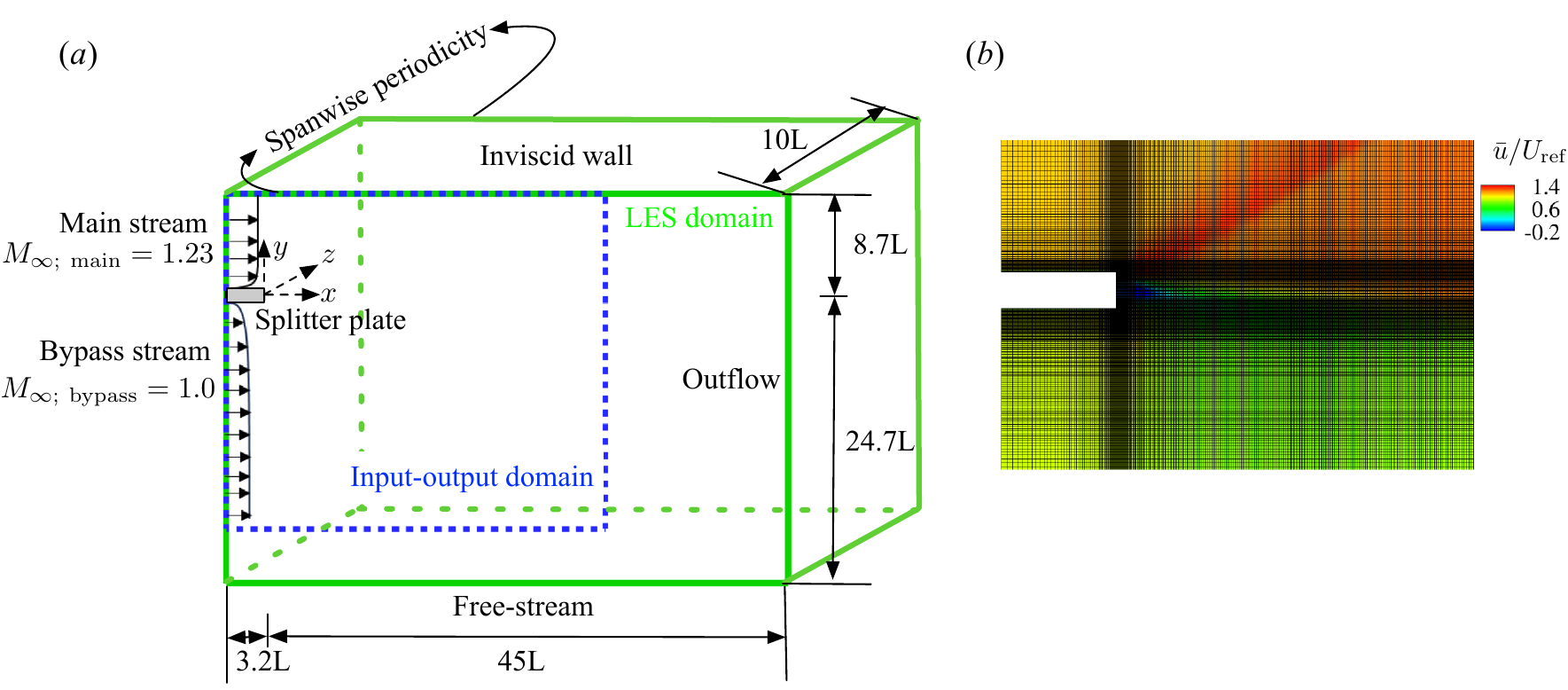}
    \caption{(a) Computational domain of the LES and a cropped domain configuration in input-output analysis, (b) The zoomed-in $x-y$ slice of time-averaged streamwise velocity ($\overline{u}/U_{ref}$) interpolated on a relatively coarse grid used for input-output and stability analyses.}
    \label{fig:domain}
\end{figure}

The numerical domain is shown in figure~\ref{fig:domain}(a). The origin of the Cartesian coordinate system is located at the center of the splitter plate trailing surface. The streamwise ($u$), transverse ($v$), and spanwise ($w$) velocities are in $x-, y-$ and $z-$ directions, respectively. Density, pressure, and temperature are denoted by $\rho$, $P$, and $T$, respectively. The spanwise-uniform free-stream Mach numbers are 1.23 and 1 for the main and bypass streams, respectively, and additional details are listed in table~\ref{tab:1}. The Reynolds number is defined as $Re = \rho_{ref} U_{ref} L / \mu_{ref} = 85{,}686$, where $L = 3.175$ mm is the splitter plate thickness, $U_{ref}$ (reference velocity) is the averaged free-stream velocity of the main and bypass streams, $\rho_{ref}$ (reference density) is the averaged free-stream density of the two streams, and the reference molecular viscosity $\mu_{ref}$ is obtained using Sutherland's law with a temperature based on the averaged free-stream temperatures. The velocity components are non-dimensionalized by the reference velocity ($U_{ref}$; the length scale (and $x-y$ coordinate systems) is non-dimensionalized by $L$; and pressure is non-dimensionalized by $P_{ref} = \rho_{ref} U_{ref}^2$. Frequencies are reported as non-dimensional Strouhal numbers, $St = f L/U_{ref}$, where $f$ is the dimensional frequency. 

\begin{table}
\begin{small}
\centering
\begin{tabular}{l l l l l l l} 
 \hline
  &  $u_\infty$[m/s] & ~~$\rho_\infty$[kg/m$^3$] & ~~$P_\infty$[Pa] & ~~$M_\infty$ & ~~$\delta_{99}$[m] & ~~$Re_{\delta_{99}}$ \\ [0.5ex] 
 \hline\hline
 Main stream & 373.47 & 2.78  & 184,323  & 1.23 & $4.51 \times 10^{-4}$  & 31,298 \\ 
 Bypass stream & 319.05 & 1.42  & 101,456  & 1.00 & $2.72 \times 10^{-4}$
 & 7,705  \\
 \hline
\end{tabular}
\caption{Freestream conditions for the main and bypass streams of the supersonic shear-layer flow.}
\label{tab:1}
\end{small}
\end{table}

In the LES simulation, the outflow and bottom boundary are placed far away from the splitter plate at distances of $45L$ and $24.7L$, respectively, to minimize the boundary effects on the shear-layer region (figure~\ref{fig:domain}(a)). The top boundary of the numerical domain is prescribed as an inviscid wall to approximate the full nozzle configuration without resolving the adjacent boundary layer. The splitter plate width is taken to be ten times its thickness and spanwise periodic boundary conditions are imposed. In figure~\ref{fig:domain}(b), a zoomed-in view $x-y$ plane of time-averaged streamwise velocity is provided with a grid topology for input-output analysis. The reader is referred to \citet{stack2019splitter} for more details on the flow configuration, grid strategy and subgrid-scale model used to generate the base flow dataset.

\subsection{Input-output framework} \label{sec:inout}

The supersonic shear-layer flow is homogeneous in the spanwise direction and the primary coherent structures display spanwise uniformity (\citet{stack2019splitter}) as shown in figure~\ref{fig:engine}(b). This motivates the decomposition of the flow state $\boldsymbol {q}(x,y,z,t) = [\rho, u, v, w, T]$ into a time-averaged two-dimensional (2-D) mean state $\boldsymbol{Q}(x,y)$ and a time-variant perturbation $\boldsymbol{q}'(x,y,z,t)$ as
\begin{equation}
    \boldsymbol{q}(x,y,z,t)=\boldsymbol{Q}(x,y)+\boldsymbol{q}'(x,y,z,t).
    \label{eq:decomposition}
\end{equation}
Strictly speaking, linearization of the Navier--Stokes (N-S) equation using the above decomposition is based on the base state $\boldsymbol Q$ being in equilibrium (\citet{trefethen1993, jovanovic2005, trefethen2005}) i.e., $\boldsymbol Q$ should be a steady solution of the N-S equations.  However, for flows at high Reynolds numbers, the statistically stationary mean (i.e., time-averaged) state is commonly used for N-S linearization, which serves as a foundation for the input-output analysis of fluid flow (\citet{liu2018, sun2017, schmidt2017, yeh2019, doshi2021}). In like manner, in the present work, we use the statistically stationary time-averaged shear-layer flow (mean flow) as the base state for the input-output analysis (\S~\ref{sec:resl}). 

The governing equation of the perturbation $\boldsymbol{q}'$ after substituting Eq.~(\ref{eq:decomposition}) into the compressible Navier--Stokes equations is 
\begin{equation}
    \frac{d {\boldsymbol q'}}{dt} = \mathcal{\boldsymbol L}({\boldsymbol Q}){\boldsymbol q'} + \boldsymbol {h'},
    \label{eq:IO1}
\end{equation}
where $ \mathcal{\boldsymbol L} (\boldsymbol Q)$ denotes the linearized compressible N-S operator about the mean state $\boldsymbol Q$. Here, $\boldsymbol h'$ is considered the nonlinear term of perturbation (\citet{mckeon2010}) as well as an external forcing. To formulate the dynamical model (Eq.~\ref{eq:IO1}) into an input-output framework, we can add an input variable ${\boldsymbol f'}$ and an output variable $\boldsymbol{y_o}$ as
\begin{equation}
\begin{split} 
   \boldsymbol{h'} = \boldsymbol{B f'}, \\
   \boldsymbol{y_o} = \boldsymbol{C q'}.
   \label{eq:IO2}
\end{split}
\end{equation}

Next, the perturbations ${\boldsymbol q}'$ and forcing ${\boldsymbol f}'$  are expressed as Fourier modes with 
\begin{equation}
\begin{split}
   \boldsymbol{q}' (x,y,z,t) =  \hat{\boldsymbol q}(x,y) e^{i(\beta z - \omega t)}, \\
   \boldsymbol{y_o} (x,y,z,t) =  \hat{\boldsymbol y_o}(x,y) e^{i(\beta z - \omega t)}, \\
   \boldsymbol{f}' (x,y,z,t) =  \hat{\boldsymbol f}(x,y) e^{i(\beta z - \omega t)}.
    \label{eq:fourier}
\end{split}
\end{equation}
where $\hat{\boldsymbol q}(x,y)$, $\hat{\boldsymbol y_o}(x,y)$ and $\hat{\boldsymbol f}(x,y)$ denote the 2-D amplitude functions of the state perturbation, output variable and the input variable (i.e., forcing), respectively; $\beta$ is a real-valued spanwise wavenumber, and $\omega$ is a complex-valued frequency. After utilizing the Fourier representation of $\boldsymbol{q}'$, $\boldsymbol{y_o}$ and $\boldsymbol{f}'$ (Eq.~(\ref{eq:fourier})), the above system of equations yield 
\begin{equation}
   \hat{\boldsymbol y_o}  = \boldsymbol{C}[ -i \omega \boldsymbol{I} - \boldsymbol{L}(\boldsymbol {Q}; \beta)]^{-1} \boldsymbol{B} \hat{\boldsymbol f} = \Tilde{\boldsymbol{H}}(\boldsymbol {Q}; \omega,  \beta) \hat{\boldsymbol f},
\end{equation}
where $\boldsymbol{I}$ is the identity matrix and $\Tilde{\boldsymbol{H}}(\boldsymbol {Q}; \omega,  \beta) = \boldsymbol{C}[ -i \omega \boldsymbol{I} - \boldsymbol{L}(\boldsymbol {Q}; \beta)]^{-1} \boldsymbol{B} $ is referred to as the {\it input-output operator}. Here, $\Tilde{\boldsymbol H}(\boldsymbol {Q}; \omega,  \beta)$ is also a transfer function between the sustained input variable $\hat{\boldsymbol f}$ to the harmonic output variable $\hat{\boldsymbol y_o}$ associated with the real-valued spanwise wave number $\beta$ and the complex-valued frequency $\omega$  for the given mean state $\boldsymbol {Q}$.

\subsubsection{Classical resolvent analysis} \label{sec:resl} 

When the input matrix $\boldsymbol{B}$ and output matrix $\boldsymbol{C}$ are identity matrices, no special constraints, such as spatial extent or restriction to specific variables, are imposed on the input or output. The input-output operator then reduces to the classical resolvent operator $\boldsymbol{H}(\boldsymbol {Q}; \omega,  \beta) = [ -i \omega \boldsymbol{I} - \boldsymbol{L}(\boldsymbol {Q}; \beta)]^{-1} $. The resolvent analysis can be cast in the framework of singular value decomposition (SVD) of the resolvent operator to determine the harmonic forcing $\hat {\boldsymbol{f}}$ and the corresponding response $\hat {\boldsymbol{q}}$. The decomposition of the resolvent operator is given by
\begin{equation}
\boldsymbol{H}(\boldsymbol {Q}; \omega,  \beta) = \boldsymbol{\mathcal{U}_q \Sigma \mathcal{V}_f}^*,
\end{equation}
where, $\boldsymbol{\mathcal{U}_q}  = [\hat {\boldsymbol q}_1,\hat {\boldsymbol q}_2,\dots,\hat {\boldsymbol q}_k]$ is a set of left singular vectors $\hat {\boldsymbol q}_j$ denoting response modes, and $\boldsymbol{\mathcal{V}_f} = [\hat {\boldsymbol f}_1,\hat {\boldsymbol f}_2,\dots, \hat {\boldsymbol f}_k]$ contains a set of right singular vectors $\hat {\boldsymbol f}_j$ denoting the forcing modes, with the superscript $^*$ representing the Hermitian transpose. The diagonal matrix $\boldsymbol{\Sigma} = \text{diag}(\sigma_1,\sigma_2,\dots,\sigma_k)$ yields the gain, with $\sigma_k^2$ representing the amplification ratio of the response and forcing modes, depending on the norm specified. The singular values are arranged in descending order $(\sigma_1\geq\sigma_2\geq...\geq\sigma_k)$, and the first singular value $\sigma_1$ is referred to as the optimal resolvent gain. If $\sigma_1 \gg \sigma_2$, a rank-1 assumption can often be appropriately made, meaning that the input-output process is dominated by the leading forcing-response pair (\citet{beneddine2016, gomez2016, towne2018, yeh2019}). The above expression can also be rewritten in terms of $\boldsymbol{H \mathcal{V}_f} = \boldsymbol{\mathcal{U}_q \Sigma}$, which facilitates interpretation of each column of $\boldsymbol{\mathcal{V}_f}$ as an input vector that is mapped into the corresponding column of the output $\boldsymbol{\mathcal{U}_q}$ through the transfer function $\boldsymbol{H}$ (\citet{schmid2002}).

In the present work, the resolvent gain is studied in the context of Chu's energy norm (\citet{chu1965}) of perturbations given by 
\begin{equation}
E = \int_\Omega \left[  \frac{R \overline{T}}{\overline{\rho}} \rho'^2 +  \overline{\rho} u_i' u_i' + \frac{R \overline{\rho}}{(\gamma-1)\overline{T}} T'^2 \right]  \,d\Omega,
\label{eq:E_chu}
\end{equation}
where $R$ is the gas constant, $\Omega$ is the entire domain used in the resolvent analysis. $\overline{(.)}$ and $(.)'$ indicate the time-averaged and the perturbation states of the variables, respectively. The norm can be related to the induced 2-norm of the state vectors through a weight matrix $\boldsymbol{W}$, such that $||\boldsymbol{\hat{q}}||^2_E = ||\boldsymbol{W\hat{q}}||^2_2 $ (\citet{schmid2002}). The weight matrix $\boldsymbol{W}$ can be constructed based on the discretization scheme adopted in the numerical configuration. We obtain the optimal ratio of Chu's energy norm of response to forcing modes via calculation of the largest singular value of the resolvent operator, which is the induced 2-norm of the weighted resolvent matrix $\boldsymbol{W}^{1/2} \boldsymbol{{H}} \boldsymbol{W}^{-1/2}$. Since we use the weight matrix $\boldsymbol{W}$ to induce the 2-norm evaluation, the resulting forcing and response modes shown later are scaled by $\boldsymbol{W}^{-1/2}$ to present the correct flowfield. 

The SVD inherent to the procedures used in this work can become prohibitively expensive in both memory and number of calculations when the input-output operator is of large size. To efficiently capture the dominant modes representing the prevailing coherent structures and dynamics, we leverage the randomized algorithm described in~\citet{ribeiro2020} to perform the SVD of the input-output operator. The algorithm and its comparison with full SVD are briefly described in appendix~\ref{sec:rsvd_method}.
  
\subsubsection{Input-output analysis configuration} \label{sec:inout_conf}

The mapping of specific inputs to corresponding outputs is performed by manipulating $\boldsymbol{B}$ and $\boldsymbol{C}$ matrices, thus providing the response of the flow to specific inputs that mimic the physical process of active flow control. This type of analysis is referred to as a componentwise input-output analysis (\citet{jovanovic2005}).  

The input variables and location are motivated by challenges and constraints in active flow control designs. In practice, the actuator can only be placed on a solid surface with specific forcing input such as a unidirectional momentum jet or thermal or acoustic wave (\citet{glezer2002, cattafesta2011, little2019, heidt2021}). Moreover, the pressure is observed in the output as the pressure fluctuations are responsible for loading on an aircraft structure and generating far-field noise.  
This motivates the use of various input variables such as $x-$, $y-$direction momentum forcing, and pressure while the output variable is the pressure. Thus, each input variable individually maps to the output pressure. In addition to the input variable, spatial restriction is also imposed on the inputs to a square box of the splitter plate thickness ($L$) near the trailing edge. Three locations, upper surface (US), trailing surface (TS), and lower surface (LS), are selected as input locations. The various state variable and location combinations can be achieved through the input matrix $\boldsymbol{B}$. No spatial restriction is imposed on the output. 

Similar to the resolvent analysis, an SVD is performed on the operator $\Tilde{\boldsymbol{H}}(\boldsymbol {Q}; \omega,  \beta)$ to obtain the output modes  $\hat{\boldsymbol y_o}$ and the input modes ${\boldsymbol{\hat f}}$. Because not all the flow state variables are considered, the amplification (gain) is studied in the context of $L2$-norm for the input-output analysis.

\subsubsection{Linear Operator Construction}

To construct the linear operator $\boldsymbol {L}(\boldsymbol Q;\beta)$ used in the classical resolvent and input-output analysis, the time-averaged base flow states on the center $x-y$ plane from the 3-D domain are used. We only consider $\beta = 0$ (i.e., 2-D modes) for the current work, and, furthermore, to reduce the computational expense while focusing on capturing coherent structures perform the analysis on a smaller domain, as shown in figure~\ref{fig:domain}(a), whose values are obtained by interpolation from the finely resolved LES. The grid used in the input-output analysis is refined in the shear layer region and stretched gradually downstream. The discretized linearized Navier-Stokes equations (\citet{sun2017}) generate the operator $\boldsymbol {L}(\boldsymbol Q;\beta=0)$ using the interpolated base flow on the coarse grid. The inlet perturbation variables for both main and bypass streams are prescribed with zero Dirichlet boundary conditions for the density, velocity components, and normal pressure gradients. On the splitter plate surfaces, all the velocity components, density, and pressure normal gradients are set to zero. For top, bottom, and outflow boundaries, zero Neumann boundary conditions are applied for density, velocity, and pressure. In addition, sponge zones are applied along the bottom and outflow boundaries to damp perturbations to weaken the influence of boundaries on the inner domain results. The top boundary is prescribed as an inviscid wall in the LES to mimic a solid wall of the nozzle. As the interaction of the reflected shock from the top boundary and shear layer is not the focus of the current work, we apply a thin sponge layer at the top boundary for the classical resolvent and input-output analyses, focusing on the perturbation dynamics near the splitter plate. Based on the grid-independence study (see appendix~\ref{sec:grid_study}), a grid with approximately 0.176 million control volumes is chosen, which yields the input-output operator in the form of a matrix with an approximate size of 0.88 million $\times$ 0.88 million. To efficiently perform the input-output analysis for a large-scale matrix, we use the randomized algorithm to compute the SVD of the input-output operator $\boldsymbol{H}(\boldsymbol {Q}; \omega)$.

\section{Results} \label{sec:result}

\subsection {Baseline flow}
\subsubsection{Instantaneous and mean flow properties} \label{sec:baseflow}

\begin{figure}
     \centering
         \includegraphics[width=0.85\textwidth]{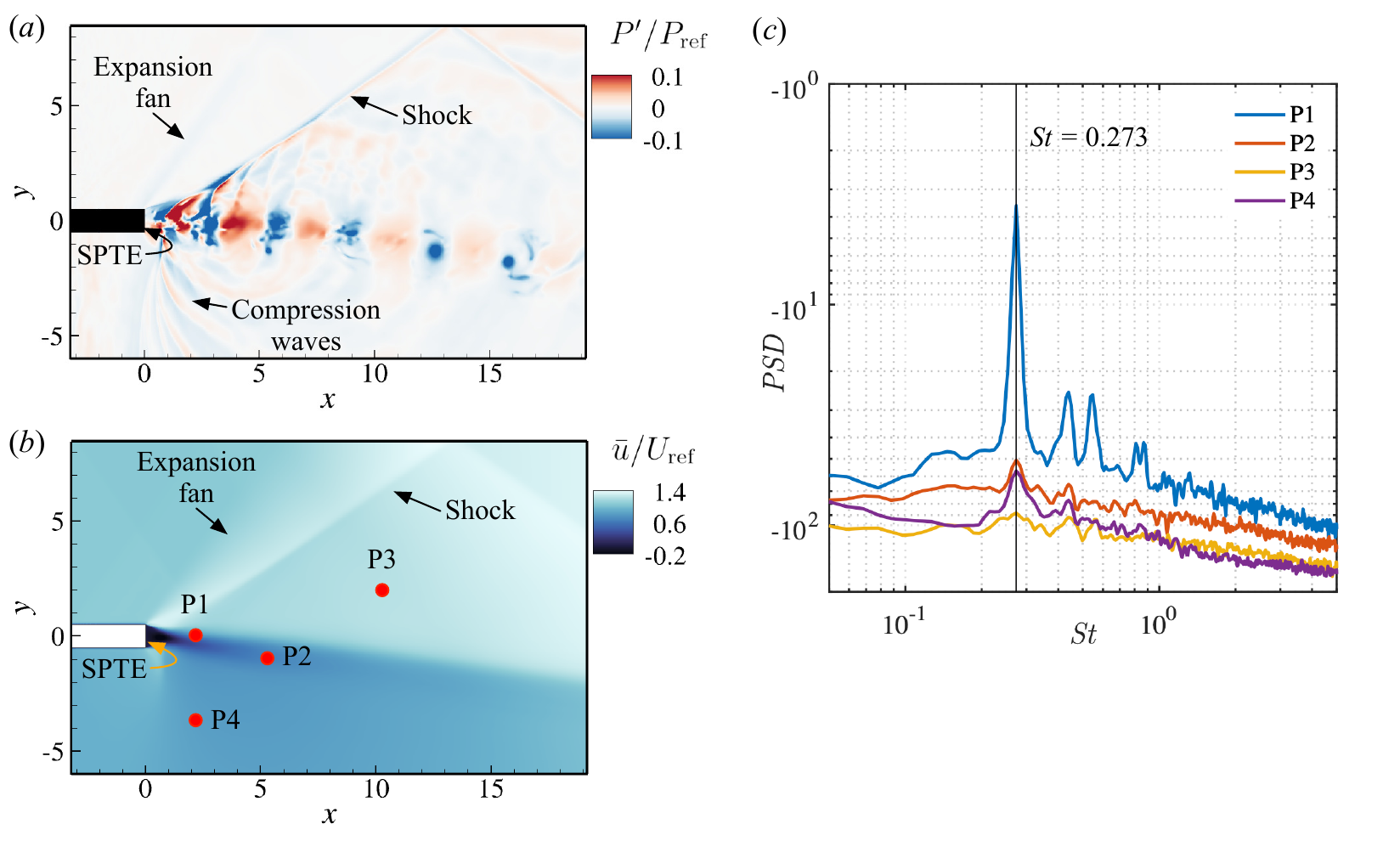}
    \caption{(a) Normalized instantaneous pressure fluctuation ($P'/P_{\text{ref}}$), (b) time-averaged streamwise velocity ($\overline{u}/U_{\text{ref}}$), and (c) power spectral density (PSD) of pressure data from probes indicated by subplot (b).}
    \label{fig:baseflow}
\end{figure}

In the baseline flow (\citet{stack2019splitter}) shown earlier in figure~\ref{fig:engine}(b), the large-scale and clockwise rotating vortical structures roll up and convect downstream along the shear layer due to the Kelvin--Helmholtz (KH) instability. For the purposes of the current work, additional features of the flow are now described. The normalized pressure fluctuation and time-averaged streamwise velocity flowfield are shown in figure~\ref{fig:baseflow}(a) and \ref{fig:baseflow}(b), respectively. A total number of 2{,}598 snapshots covering the convective time of $t U_{\text{ref}}/L \approx 260$ is used to generate the mean flow. Observed from the instantaneous and mean flow features, the supersonic main stream expands at the corner of the SPTE due to the downward deflection, which results in an expansion fan on the upper side of SPTE. The supersonic main stream then interacts with the sonic bypass stream (near probe P1), and an oblique shock redirects the flow back to the streamwise direction. This oblique shock propagates upward and reflects from the top boundary of the domain and interacts with the shear layer further downstream, far away from the SPTE. We note that the region of interaction between this reflected shock and the shear layer is not included in the domain visualized here. Moreover, a recirculation zone is observed in the vicinity of the SPTE, similar to a wake flow behind a bluff body and the two streams form a shear layer as the flow mixes further downstream. On the lower side of the splitter plate, the bypass stream also experiences a slight expansion due to the sudden increase in the flow area at the SPTE, but no distinct shock is apparent in the mean flowfield. Meanwhile, the weaker compression waves propagating into the bypass stream from SPTE are observed in the unsteady fluctuating pressure $P'$, as shown in figure~\ref{fig:baseflow}(a). 

The unsteadiness of the vortical structures is analyzed using data from four probes (indicated by red dots in figure~\ref{fig:baseflow}(b)) placed at representative locations in the flowfield. The power spectral density (PSD), calculated using the time-history of pressure, is shown in figure~\ref{fig:baseflow}(c). Probes P1 and P2 are located in the shear layer downstream of the SPTE, while P3 and P4 are located on the upper and lower sides of the shear layer, respectively. As shown in figure~\ref{fig:baseflow}(c), a distinct peak is observed at $St=0.273$ for all the probes, and the pressure fluctuations in the shear layer (i.e., P1 and P2) are larger than those on the upper (P3) and lower (P4) sides of the shear layer. Moreover, the fact that all probes capture the same peak frequency, indicates that the far-field acoustics (\citet{magstadt2015, berry2016}) is highly influenced by the formation of the vortical structures after the splitter plate.

\subsubsection {Spectral proper orthogonal decomposition} \label{sec:SPOD}
 
Spectral analysis methods can identify coherent structures associated with each frequency in turbulent flows (\citet{lumley1967, Rowley2009, Sieber2016, towne2018}). In the present work, we use the spectral proper orthogonal decomposition (SPOD) algorithm discussed in detail by (\citet{towne2018, schmidt2020}). Modes calculated from this method are orthogonal to other modes at the same frequency and optimally represent spatio-temporal flow statistics. 

To investigate coherent structures of the most energetic modes in the supersonic shear layer, the SPOD analysis is conducted on the centerplane $x-y$ flowfield data, interpolated to a relatively uniform coarse grid. A rectangular domain $(-3.2 \le x \le 30, -10 \le y \le 8.7)$ discretized by approximately 0.25 million grid points is used for the interpolation and the state variables $[\rho', u', v', w', T']$ are used to perform SPOD. 2{,}598 snapshots spaced at equal time intervals of $\triangle t U_\text{ref} / L = 0.1$ are used. We perform the SPOD analysis for 128, 256, 512, and 1024 snapshots per block with $50\%$ overlap to capture the dominant physics. Based on the convergence of the modal structures and frequency resolution, we select 512 snapshots per block with $50\%$ overlap, resulting in 9 blocks, for further discussion.

\begin{figure}
     \centering
         \includegraphics[width=1\textwidth]{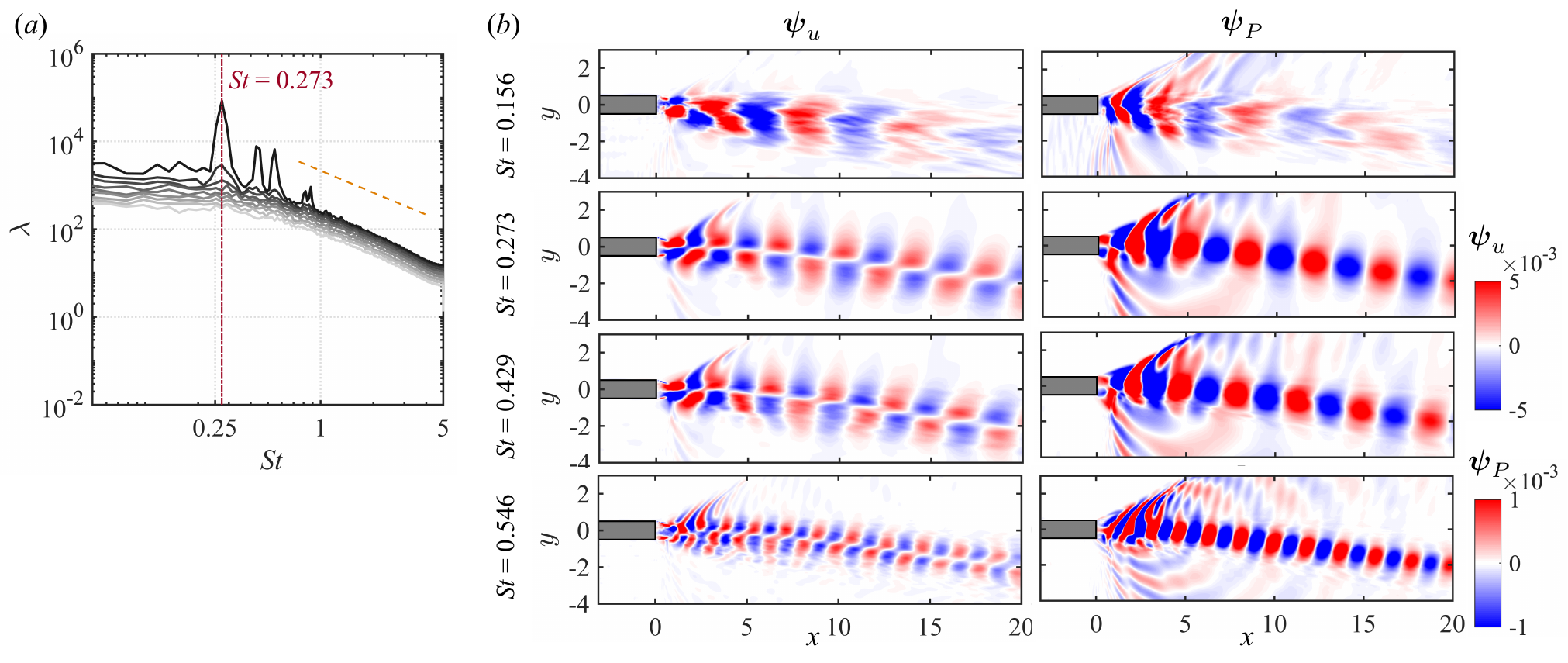}
    \caption{(a) SPOD eigenvalue spectrum of the supersonic shear layer. Decreasing eigenvalues are shown in lighter shades, i.e., $\lambda_1  \geq \lambda_2 \geq \dots \geq \lambda_N$ at each frequency. The dashed orange line indicates the $-5/3$ power law of the energy spectrum. (b) The leading SPOD mode (first mode) of the streamwise velocity $\psi_u$ and the pressure $\psi_P$ over a frequency range from $St = 0.156$ to 0.546.}
    \label{fig:spod}
\end{figure}

In SPOD analysis, the eigenvalues represent the energy of the modal structures based on a norm weight. In this study, Chu's energy norm (Eq.~\ref{eq:E_chu}) is considered. The SPOD eigenvalue spectrum of the supersonic shear layer flow is shown in figure~\ref{fig:spod}(a). The peak in the eigenspectra is observed at $St = 0.273$ for the leading mode ($\lambda_1$), which is significantly larger than the second mode energy ($\lambda_2$). This indicates that the leading mode is much more energetic than the other modes, and the primary physical mechanism is dominated by the leading mode. In other words, the flow exhibits a low-rank behavior at frequency $St=0.273$. The low-rank mechanism is also observed at the frequencies $St=0.117,0.156$, 0.429, along with the first harmonic of 0.429 and the first two harmonics of the dominant frequency.
All these peak frequencies are also captured in the spectral analysis of the baseline flow probes data (see figure~\ref{fig:baseflow}(c)).

Figure~\ref{fig:spod}(b) shows the streamwise velocity $\psi_u$ and pressure $\psi_P$ leading SPOD modes at four representative frequencies. The pressure modes are obtained by using a linearized ideal gas law,
\begin{equation}
\psi_P = R(\overline{\rho} \psi_T + \overline{T} \psi_\rho).
\end{equation}
The streamwise velocity field is located in the shear-layer region with alternative positive and negative lobe structures for the dominant frequency $St=0.273$. These structures resemble the Kelvin-Helmholtz shear-layer instability of the mean flow. The pressure mode takes the form of the wavepacket in the shear-layer region with far-field traveling acoustic waves along the oblique shock and the compression waves in the upper and lower stream, respectively. For the other high-frequency peaks at $St = 0.429$, and $0.546$, the velocity and pressure mode structures are similar to those of the dominant frequency but with a smaller wavelength along the streamwise direction. The far-field traveling waves are also observed at these frequencies, revealing the global spatiotemporal consequences of the supersonic shear layer. The mode structures at low frequency $St = 0.156$ do not exhibit clear wavepacket structures in the layer; these highly distorted structures may be affected by diminished convergence of low-frequency statistics. The eigenspectra and modes obtained from SPOD are compared with the optimal gain and modes of classical resolvent analysis in \S~\ref{sec:res_SPOD}.

\subsection{Classical resolvent analysis}
\subsubsection{2-D instabilities of the base flow} \label{sec:stab}

\begin{figure}
     \centering
         \includegraphics[width=0.75\textwidth]{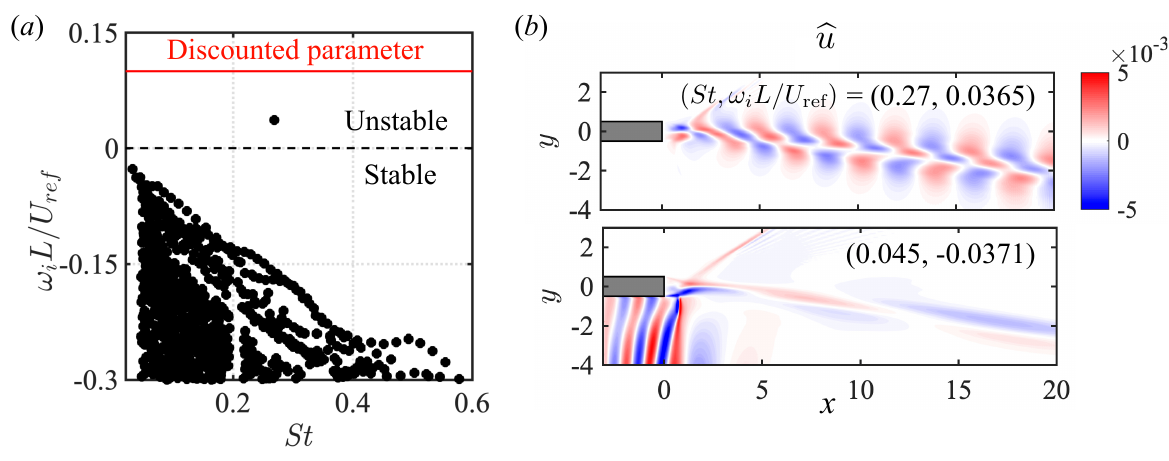}
    \caption{2-D eigenmodes of stability analysis. (a) Eigenvalues of $\boldsymbol {L}(\boldsymbol {Q}; \beta)$ with $\beta=0$. The red line indicates the discounted parameter value. (b) Modal structures of two representative eigenmodes, the real component of $\hat{u}$.}
    \label{fig:stability}
\end{figure}

A 2-D stability analysis is first performed on the time-averaged mean flow (\citet{barkley2006, turton2015, beneddine2016, sun2017, Sun:AIAAJ19_2}). The eigenvalues of the linear operator $\boldsymbol {L}(\boldsymbol {Q})$ are shown in figure~\ref{fig:stability}(a). The frequency $\omega_r$ (the real component of the eigenvalue $\omega$) is reported using a normalized Strouhal number as $St = \omega_r L / (2 \pi U_{\text{ref}}$), and the growth/decay rate $\omega_i$ (the imaginary component of the eigenvalue $\omega$) is normalized as $\omega_i L / U_{\text{ref}}$. The black dashed line separates the stable and unstable regions of the complex plane. The linear operator $\boldsymbol {L}(\boldsymbol {Q})$ is unstable as the leading eigenvalue is with a positive growth rate of $\omega_i L / U_{\text{ref}} = 0.0365$ at $St = 0.27$. The frequency of this most unstable 2-D eigenmode agrees well with the peak frequency captured in the power spectra density and SPOD analyses. 

The modal structures of the streamwise velocity $\hat{u}$ for the two representative dominant eigenmodes are shown in figure~\ref{fig:stability}(b). The most unstable eigenmode with eigenvalue of $(St, \omega_i L / U_{\text{ref}}) = (0.27, 0.0365)$ presents a distinctive structure distributed along the shear layer, again resembling the pattern of Kelvin--Helmholtz vortex street, whereas the sub-dominant (least stable) mode with eigenvalue of $(St, \omega_i L / U_{\text{ref}}) = (0.045, -0.0371)$ displays elongated structures in the shear layer as well as upstream propagating waves on the lower side of the splitter plate. Comparing the eigenmode structures in the shear layer, the lower-frequency mode has a larger streamwise wavelength. The spurious eigenmodes that are associated with unphysical flow patterns (\citet{lesshafft2018}) are not discussed here. 

\subsubsection{Resolvent spectra and modes} \label{sec:resl_result} 

The classical resolvent analysis is performed for the full-state variables $[\rho', u', v', w', T']$ by using the time-averaged mean data from the center $x-y$ plane. As indicated by the eigenvalue spectrum from the stability analysis (figure~\ref{fig:stability}), the linear operator $\boldsymbol {L}(\boldsymbol {Q})$ built about the turbulent mean flow is unstable. Hence, the resolvent analysis should be modified by incorporating a real-valued parameter $\alpha$ such that $\alpha > \text{max}(\omega_i) $. This discounted resolvent analysis was proposed by (\citet{jovanovic2004}) to analyze unstable dynamical systems. To avoid unbounded energy amplification in performing the resolvent analysis, the original resolvent operator $\boldsymbol{H}(\boldsymbol {Q}; \omega) = [ -i \omega \boldsymbol{I} - \boldsymbol{L}(\boldsymbol {Q})]^{-1} $ is then modified as (\citet{Sun:AIAAJ19_2})      
\begin{equation}
\boldsymbol{H}^\alpha (\boldsymbol {Q}; \omega) = [ -i (\omega + i\alpha) \boldsymbol{I} - \boldsymbol{L}(\boldsymbol {Q})]^{-1} = [ -i \omega \boldsymbol{I} - (\boldsymbol{L}(\boldsymbol {Q}) -\alpha \boldsymbol{I})]^{-1}.
    \label{eq:dis_resol}
\end{equation}
The expression provides the discounted resolvent operator $\boldsymbol{H}^\alpha (\boldsymbol {Q}; \omega)$ by shifting the linear operator as $(\boldsymbol{L}(\boldsymbol {Q}) -\alpha \boldsymbol{I})$; in other words, the eigenspectrum of $\boldsymbol {L}(\boldsymbol {Q})$ is now shifted by $-\alpha$ along the imaginary axis, and all eigenvalues of the shifted linear operator lie on the stable plane. An equivalent exercise may be performed by directly evaluating the pseudospectrum of $\boldsymbol {L}(\boldsymbol {Q})$ on a raised frequency axis of $\omega_r + i\alpha$. Thus, $\alpha$ may also be viewed as a finite-time window, characterized by $1/ \alpha$, to investigate the flow response. For $\alpha \rightarrow 0^+$, the time window becomes infinite such that Eq.~(\ref{eq:dis_resol}) reduces to the original resolvent operator $\boldsymbol{H}(\boldsymbol {Q}; \omega)$. Because the most unstable mode of $\boldsymbol {L}(\boldsymbol {Q})$ has a normalized growth rate of $\omega_i L/U_\text{ref} = 0.0365$, the resolvent computation is carried over a temporal window characterized by a discounted parameter of $\alpha L/U_\text{ref} = 0.1$ added to the original formulation of the resolvent operator as given in Eq.~(\ref{eq:dis_resol}). The imaginary frequency axis about which the resolvent analysis is carried out is shown by the red line in figure~\ref{fig:stability}(a). Appendix~\ref{sec:dis_effect} demonstrates that the discounted parameter does not affect the amplification mechanism, thereby justifying the use of the discounted resolvent in the present study. The SVD is performed on the discounted resolvent operator $\boldsymbol{H}^\alpha (\boldsymbol {Q}; \omega)$ for the following discussion.

\begin{figure}
     \centering
         \includegraphics[width=0.4\textwidth]{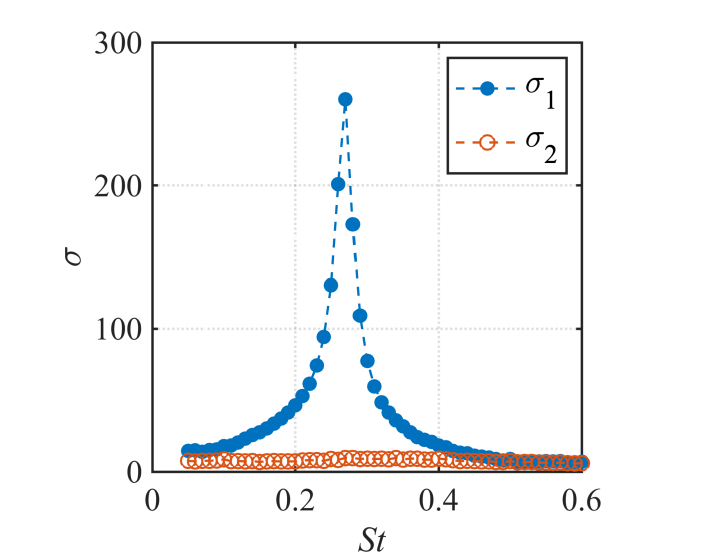}
    \caption{First ($\sigma_1$) and second ($\sigma_2$) singular values of the resolvent analysis using Chu-norm-based gain for full state variables without any spatial window.}
    \label{fig:gain_chu}
\end{figure}

The first ($\sigma_1$) and second ($\sigma_2$) singular values (gain) are shown in figure~\ref{fig:gain_chu}. The qualitative trend of the optimal gain is similar to past studies carried out to the same shear layer by \citet{doshi2022} to motivate a passive control strategy, though the values are not equivalent due to the use of different discounted parameter values and state variables (\citet{karban2020}). The highest gain ($\sigma_1$) is observed at $St = 0.27$, which is the vortex roll-up frequency in the shear layer observed in the baseline flow. $\sigma_1$ decreases sharply at frequencies away from $St = 0.27$, indicating that the dominant mechanism of energy amplification is the KH instability in the shear layer. The sub-optimal gain ($\sigma_2$) shows approximately an order of magnitude lower amplification compared to the optimal gain ($\sigma_1$) around the dominant frequency of $St = 0.27$. Hence, the flow exhibits a rank-1 behavior at the dominant frequency. The $\sigma_1$ and $\sigma_2$ are comparable at the lower frequencies ($St \lesssim 0.1$) and higher frequencies ($St \gtrsim 0.4$). In these frequency ranges, the rank-1 behavior is not valid.

\begin{figure}
     \centering
         \includegraphics[width=1\textwidth]{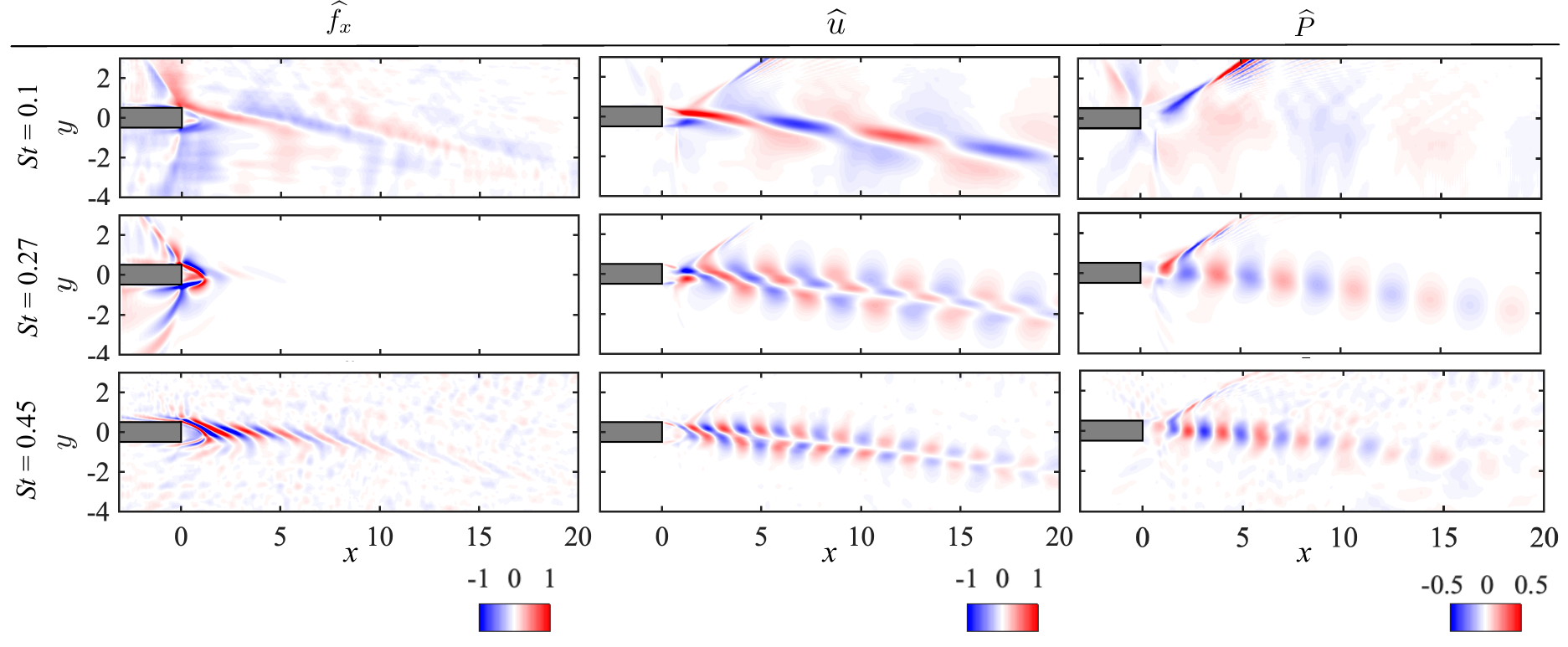}
    \caption{Spatial structures of optimal resolvent modes. From left to right, $x-$direction forcing modes ($\hat{f}_x$), response mode of streamwise velocity ($\hat{u}$) and response mode of pressure perturbation ($\hat{P}$) at (top to bottom) three different frequencies of $St = 0.1, 0.27$ and 0.45.}
    \label{fig:modes_chu}
\end{figure}

The spatial structures of the optimal resolvent modes (forcing and response modes) at three different frequencies, $St = 0.1, 0.27$, and $0.45$, are shown in figure~\ref{fig:modes_chu}. Here, we only focus on the first optimal mode as the suboptimal modes are not fully resolved. The pressure mode is obtained using the same linearized gas law as described above. The streamwise velocity $\hat{u}$ and pressure $\hat{P}$ response modes show that the disturbance propagates primarily along the shear layer and the oblique shock region close to the trailing edge. As the frequency increases, the number of lobes per unit distance along the shear layer increases because smaller wavelengths are associated with higher frequencies. The strength of the disturbance propagating along the shock is comparable to shear-layer wavepackets for the low frequency $St=0.1$, which suggests that the shock pattern is more responsive to lower-frequency excitation. At the dominant frequency of the flow $St = 0.27$, the streamwise forcing structures ($\hat{f}_x$) are primarily located in the vicinity around the splitter plate, including its upper, lower, and trailing areas. At the lower ($St=0.1$) and higher ($St=0.45$) frequencies $St=0.1$, the forcing structure stretches further in the shear layer compared to the locally concentrated forcing of $St=0.27$. 

The type of amplification mechanism in the flow at particular frequencies can be identified by the degree of spatial overlap between the forcing and response modes (\citet{symon2018}). At the resonant frequency of $St = 0.27$, the forcing mode is centered around the SPTE, while the response mode extends downstream along the shear layer. This indicates that the convective Kelvin- Helmholtz instability is the dominant energy-amplification mechanism at $St = 0.27$. Despite this, a small region of absolute instability exists near the SPTE, where the forcing and response modes overlap, resulting in a nonzero projection. The primary amplification mechanism for $St = 0.1$ and $0.45$ is absolute instability, as there is no zero projection at any spatial location along the shear layer.

Leveraging the natural amplification mechanism of the flow identified in classical resolvent analysis poses a challenge for flow control. The forcing modes from the resolvent analysis are globally distributed in the shear layer and far from the splitter plate surface (figure~\ref{fig:modes_chu}), making it technically difficult to introduce perturbations that match the predicted forcing modes and effectively trigger flow dynamics. Given that current active control techniques can only place actuators on the surfaces of a bluff body to affect the base flow, we adopt the input-output analysis with a constrained spatial window for the input, as discussed in \S~\ref{sec:io}.  

\subsection{Comparison between the resolvent and SPOD} \label{sec:res_SPOD}

In this section, we compare the resolvent and SPOD results to facilitate the interpretation of the input-output analysis below in terms of their implication on coherent structures and implicit modulation of nonlinear forcing. This comparison is motivated by a recent theoretical connection between resolvent and SPOD (\citet{semeraro2016, towne2018}). \citet{towne2018} demonstrated that the resolvent modes are identical to the SPOD modes if the forcing is uncorrelated with equal amplitude in the entire flowfield, representing unit variance white-noise forcing. In the case of high-speed turbulent flows, nonlinear forcing terms are correlated, deviating from white-noise forcing. As a result of these correlated nonlinear forcing terms, the resolvent modes often differ from the SPOD modes at the same frequency.

\begin{figure}
     \centering
         \includegraphics[width=0.7\textwidth]{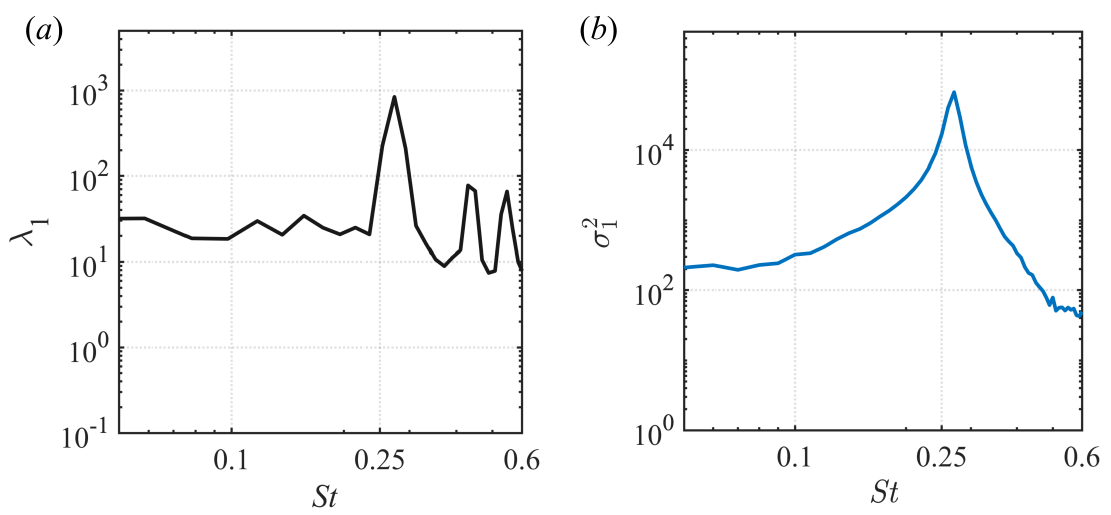}
    \caption{(a) The leading SPOD energy spectrum ($\lambda_1$) shown in zoomed-in window from figure~\ref{fig:spod}(a). (b) The optimal energy amplification ($\sigma_1^2$) is obtained through classical resolvent analysis.}
    \label{fig:gain_vs_energy}
\end{figure}

We first compare the leading energy SPOD spectrum ($\lambda_1$) and optimal resolvent energy amplification ($\sigma_1^2$) in figure~\ref{fig:gain_vs_energy}. The distribution of resolvent energy amplification closely matches that of the leading SPOD energy. Both the SPOD and resolvent spectra peak at the dominant frequency $St=0.27$, indicating low-rank behavior. However, the other peaks in the SPOD spectra at $St=0.117$, 0.156, 0.429, and 0.546 are absent in the resolvent spectrum. The nonlinear energy transfer in the shear layer may result in peaks other than $St=0.273$ in SPOD that are not well-captured by the linearized resolvent analysis. In the following discussion, we will inspect the nonlinear interactions. Additionally, the 2-D resolvent analysis may overlook the three-dimensionality inherent in the center plane. This is beyond the scope of our current work, and we focus on the dominant linear dynamics of the flow in the following section.

Next, we evaluate the mode alignment coefficient by projecting the resolvent modes onto SPOD modes to compare the resolvent and SPOD modal structures. The mode alignment coefficient $\mathscr{A} = | \psi^*_1 \boldsymbol{W} \hat{q}_1 |$ is computed between the leading SPOD mode ($\psi_1$) and optimal resolvent response mode ($\hat{q}_1$) with all state variables. The projection is computed in the spatial window $(0 \le x \le 15, -10 \le y \le 5)$ to focus on the downstream region of the SPTE with Chu's energy ($\boldsymbol{W}$). The alignment coefficient $\mathscr{A}$ ranges from 0 to 1, with perfect alignment occurring when $\mathscr{A} = 1$. Three frequencies, $St \approx 0.27, 0.43$ and 0.55 are selected based on the peaks observed in the SPOD spectra to compute the alignment coefficient. The lower frequencies are avoided as SPOD modes show distorted structures. The alignment coefficients $\mathscr{A}$ are 0.541, 0.293, and 0.155 for $St \approx 0.27$, 0.43, and 0.54, respectively. For the higher frequencies, the coefficient is lower than at the dominant frequency ($St =0.27$). From the comparison between SPOD and resolvent results, the resolvent is able to capture the dominating mechanism in the real flow at the frequency $St=0.27$.

As the resolvent spectrum is not able to capture peaks other than the KH instability $St=0.27$, we employ the bispectral mode decomposition (BMD) to uncover nonlinear physical mechanisms by examining higher-order flow statistics (for a detailed algorithm, refer to \citet{schmidt2020bmd}). The $u-$velocity is considered using the $L2$ energy norm, with the same number of snapshots per block as the SPOD for spectrum estimation.

\begin{figure}
     \centering
         \includegraphics[width=0.8\textwidth]{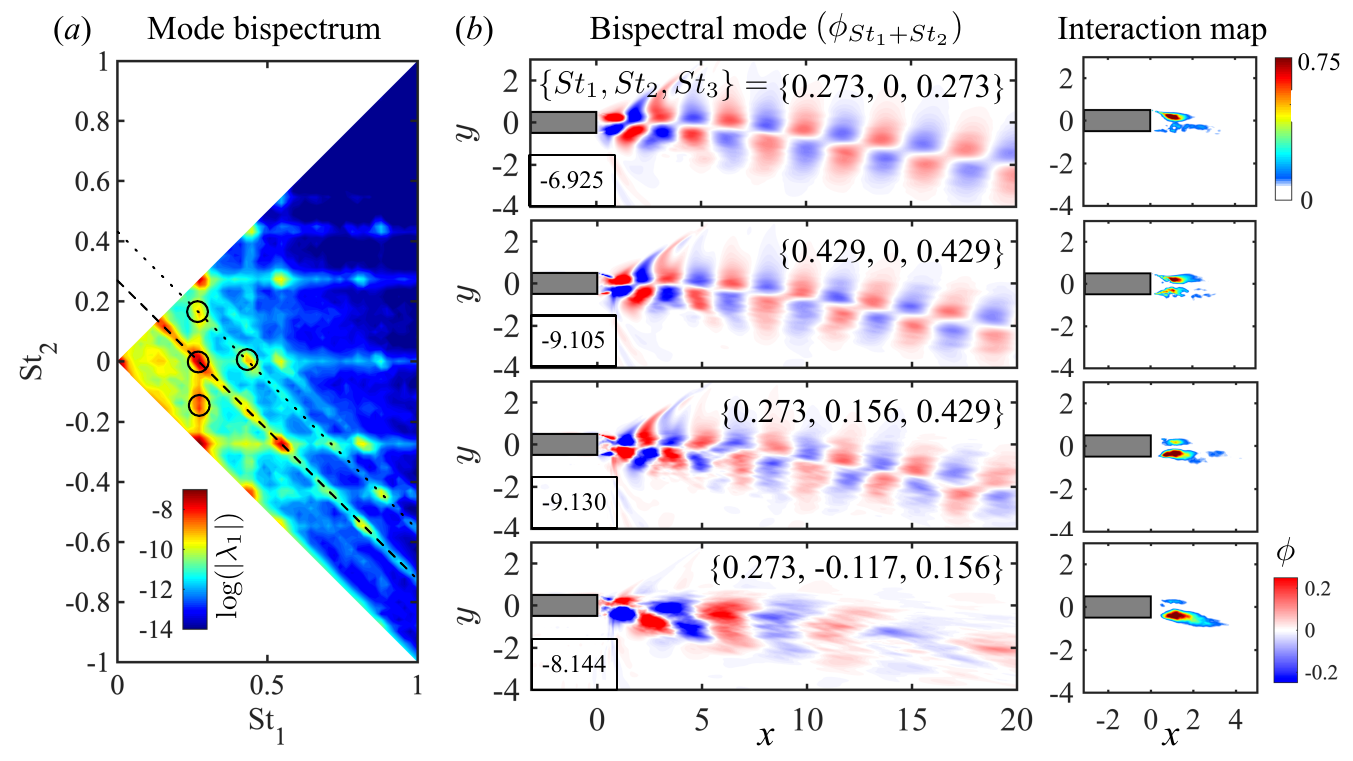}
    \caption{(a) Mode bispectrum (magnitude) in the sum and difference regions. Dash (- -) and dotted (\dots) lines represent the constant frequencies $St=0.273$ and $0.429$, respectively. (b) Bispectral modes (real part of $u-$velocity) and interaction maps for the frequency triads are marked in (a). The values in the bottom left corner of the bispectral mode display the value of $\text{log}(|\lambda_1|)$ for the corresponding triads.}
    \label{fig:bmd_mode}
\end{figure}

Figure~\ref{fig:bmd_mode}(a) displays the mode bispectrum over the regions of sum interactions ($St_1 + St_2 = St_3$) and difference interactions ($St_1 - St_2 = St_3$). The global peak occurs when a fundamental KH frequency ($St = 0.273$) interacts with itself in a difference manner to generate mean flow deformation, i.e., $\{St_1, St_2, St_3\} = \{0.273, -0.273, 0\}$. Along the $St_2 = 0$ line, which represents the linear evolution of the mean flow, local maxima are observed at $St = 0.273$, indicating the dominance of the KH instability. The first harmonic of $St = 0.273$ is primarily generated through a nonlinear energy transfer of the KH instability with itself, represented by $\{0.273, 0.273, 0.546\}$. The dotted line indicates the constant frequency ($St = 0.429$) line. Along this line, the two most dominant triads are $\{0.429, 0, 0.429\}$ and $\{0.273, 0.156, 0.429\}$. This shows that the frequency $St = 0.429$ is fed energy from the mean flow and a strong nonlinear interaction between the KH instability and $St=0.156$. We also observe other dominant triads at $\{0.273, -0.117, 0.156\}$, $\{0.273, -0.156, 0.117\}$, $\{0.156, 0.117, 0.273\},$ and $\{0.546, -0.273, 0.273\}$. 

Before examining the energy cascade among these frequencies, the bispectral mode ($\phi_{St_1 + St_2}$) and the interaction map are depicted in figure~\ref{fig:bmd_mode}(b) for the triads marked in~\ref{fig:bmd_mode}(a). The bispectral modes closely resemble the SPOD $u-$velocity modes at the same frequency.  The mode at $St=0.429$, generated by the interaction between $St=0.273$ and 0.156, displays insignificant distorted structures. Moreover, the interaction maps indicate that the dominant and higher ($St = 0.429$) frequencies interact with the mean flow on the upper side of the shear layer. In contrast, the low frequencies ($St = 0.117$ and 0.156) interact with the fundamental KH instability ($St = 0.273$) on the lower side of the shear layer.

\begin{figure}
     \centering
 \includegraphics[width=1\textwidth]{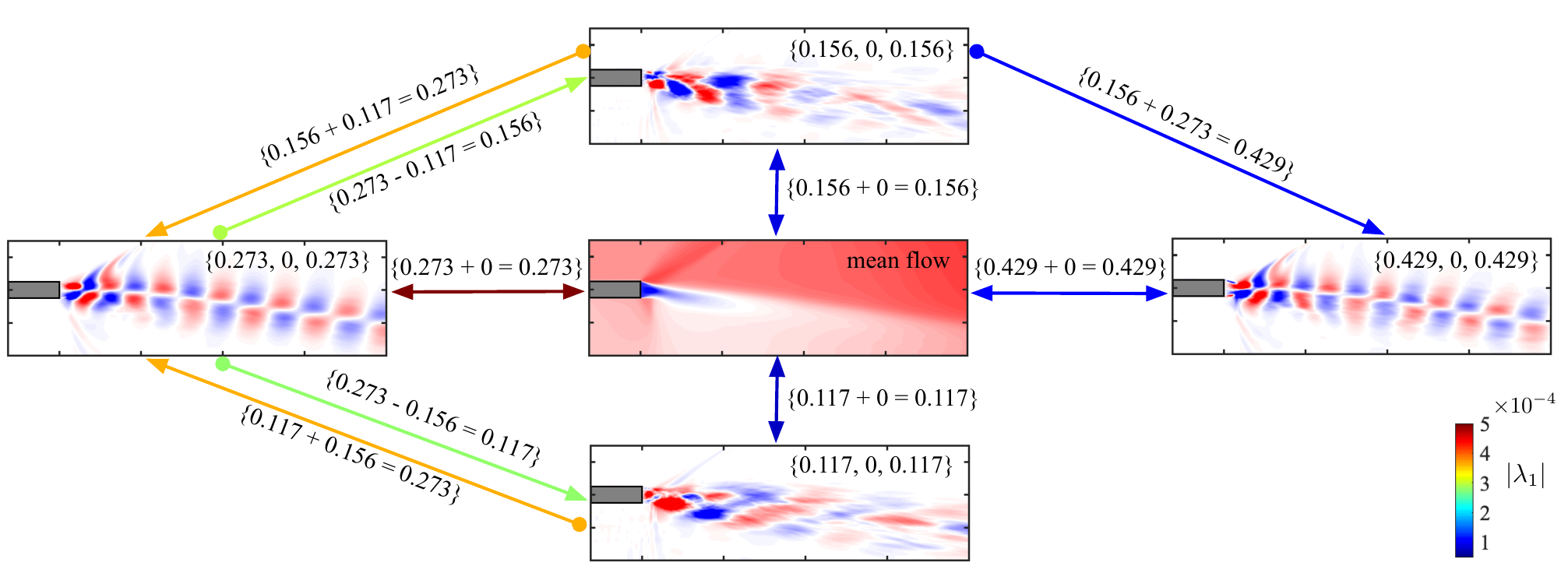}
    \caption{ The nonlinear energy cascade of dominant triads for the generation of $St=0.117, 0.156$, 0.273, and 0.429. The arrows are colored by the absolute value of mode bispectrum $|\lambda_1|$.}
    \label{fig:bmd_inter}
\end{figure}

To conclude the discussion on the nonlinear energy cascade using BMD, we plot an interaction map among frequencies that show a peak in the SPOD spectra in figure~\ref{fig:bmd_inter}. The frequencies $St=0.117$ and $St=0.156$ mainly interact with the fundamental KH instability ($St=0.273$) to produce $St = 0.156$ and 0.117, respectively. The KH instability is prominently fed energy from the mean flow. Additionally, $St=0.117$ and $0.156$ contribute to the energy cascade to $St=0.273$. The frequency $St=0.429$ receives energy almost equally from the mean flow and the nonlinear interaction between $St=0.156$ and the KH instability. BMD reveals that the frequencies $St=0.117$ and $St=0.156$ are primarily produced by nonlinear interactions, whereas both the mean flow and nonlinear interactions are almost equally responsible for generating the frequency $St=0.429$. The BMD results show that the energetic frequencies $St=0.117$, 0.156, 0.429, and 0.546 in SPOD mainly arise from the nonlinear energy cascade of the KH instability with the other frequencies or itself. Hence, it is reasonable that the linearized resolvent spectrum is unable to capture peaks at those frequencies.

\subsection{Componentwise input-output analysis} \label{sec:io}

We now investigate different combinations of forcing components as inputs and flow responses as outputs. To quantify the flow unsteadiness in the shear layer region, the pressure is examined as the output variable. Considering feasible control techniques that can introduce perturbation into the flow through a surface, suitable regions of velocity components and pressure are chosen separately as inputs to the fluid dynamical system.

\subsubsection {Velocity as an input} \label{sec:io_velocity}
We first examine the effect of input velocity forcing, deemed spatial momentum-based forcing, on output pressure fluctuations. Since the output depends on spatial restrictions of the forcing, different spatial restrictions on forcing are considered. The largest amplification ($\sigma_1^2$) from input to output is quantified using the $L2$ norm. 

\begin{figure}
     \centering
         \includegraphics[width=0.9\textwidth]{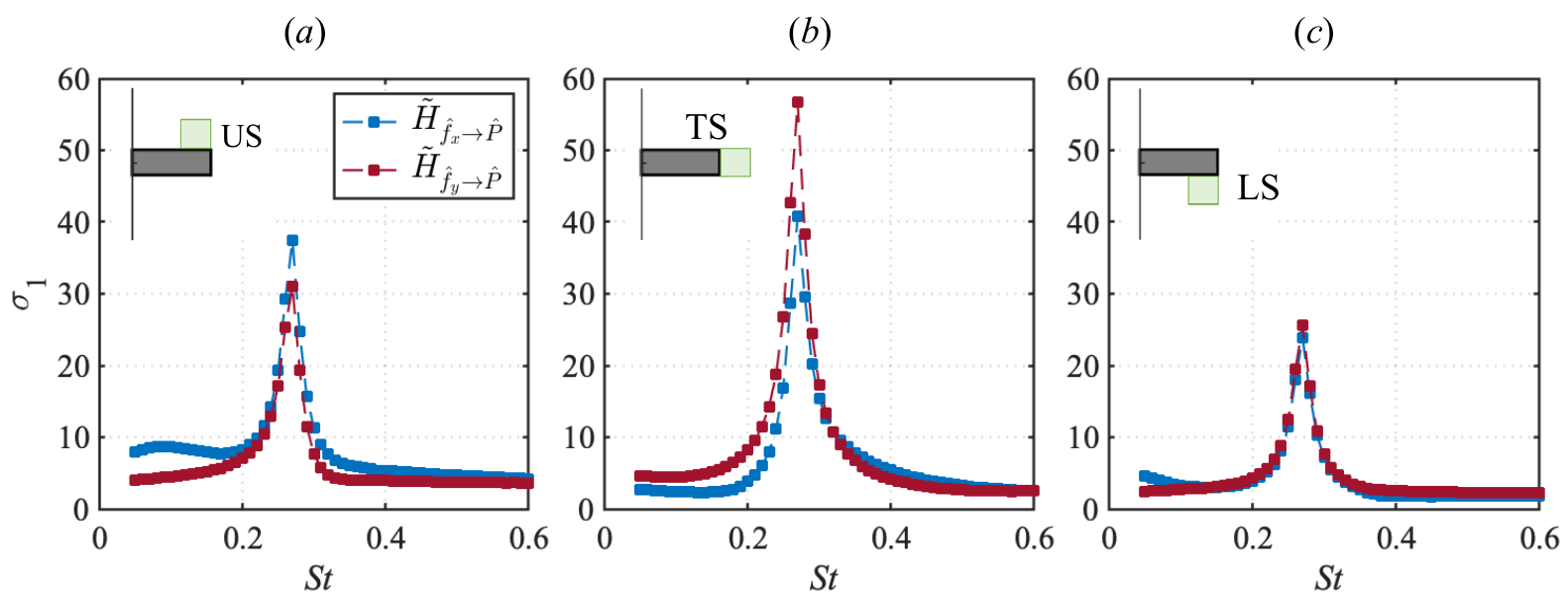}
    \caption{$L2$-norm-based gain with $x-$ and $y-$direction input variable and pressure as output variable. Input is constrained at the (a) upper surface (US), (b) trailing surface (TS), and (c) lower surface (LS) of the splitter plate. The green boxes denote an input applied area. $\tilde{\boldsymbol{H}}_{\hat{f}_x \rightarrow P}$ denotes the case with $x-$direction input while pressure is output variable.}
    \label{fig:gain_velocity}
\end{figure}

The optimal gains ($\sigma_1$) for three different locations, informed by the prior analyses, with momentum components as input variables, are shown in figure~\ref{fig:gain_velocity}. The input forcing is spatially constrained around the surfaces of the splitter plate: upper surface (US), trailing surface (TS), and lower surface (LS), as indicated by a green box in figure~\ref{fig:gain_velocity}. The highest gain is observed at the dominating KH frequency $St = 0.27$ regardless of input location and variable. The trend of $\sigma_1$ is qualitatively similar to the classical resolvent spectrum. 

When forcing is applied at different locations on the splitter plate — US, TS, and LS — the optimal input direction (either $\hat{f}_x$ or $\hat{f}_y$) varies depending on the frequency range. At lower frequencies ($0 < St \lesssim 0.15$), $\hat{f}_x$ tends to yield higher gains when introduced at the US. Conversely, $\hat{f}_y$ produces a higher gain at the TS for frequencies up to $St \lesssim 0.3$. However, for the higher frequency region ($St \gtrsim 0.3$), gains become comparable between $\hat{f}_x$ and $\hat{f}_y$ for both US and TS input. When introducing forcing at the LS (figure \ref{fig:gain_velocity}(c)), the amplifications in the pressure are almost identical regardless of the forcing direction. Across all scenarios, the TS consistently shows the highest amplification, particularly with $\hat{f}_y$, suggesting it is the optimal location for inducing significant pressure fluctuations in the flow.

\begin{figure}
     \centering
         \includegraphics[width=0.95\textwidth]{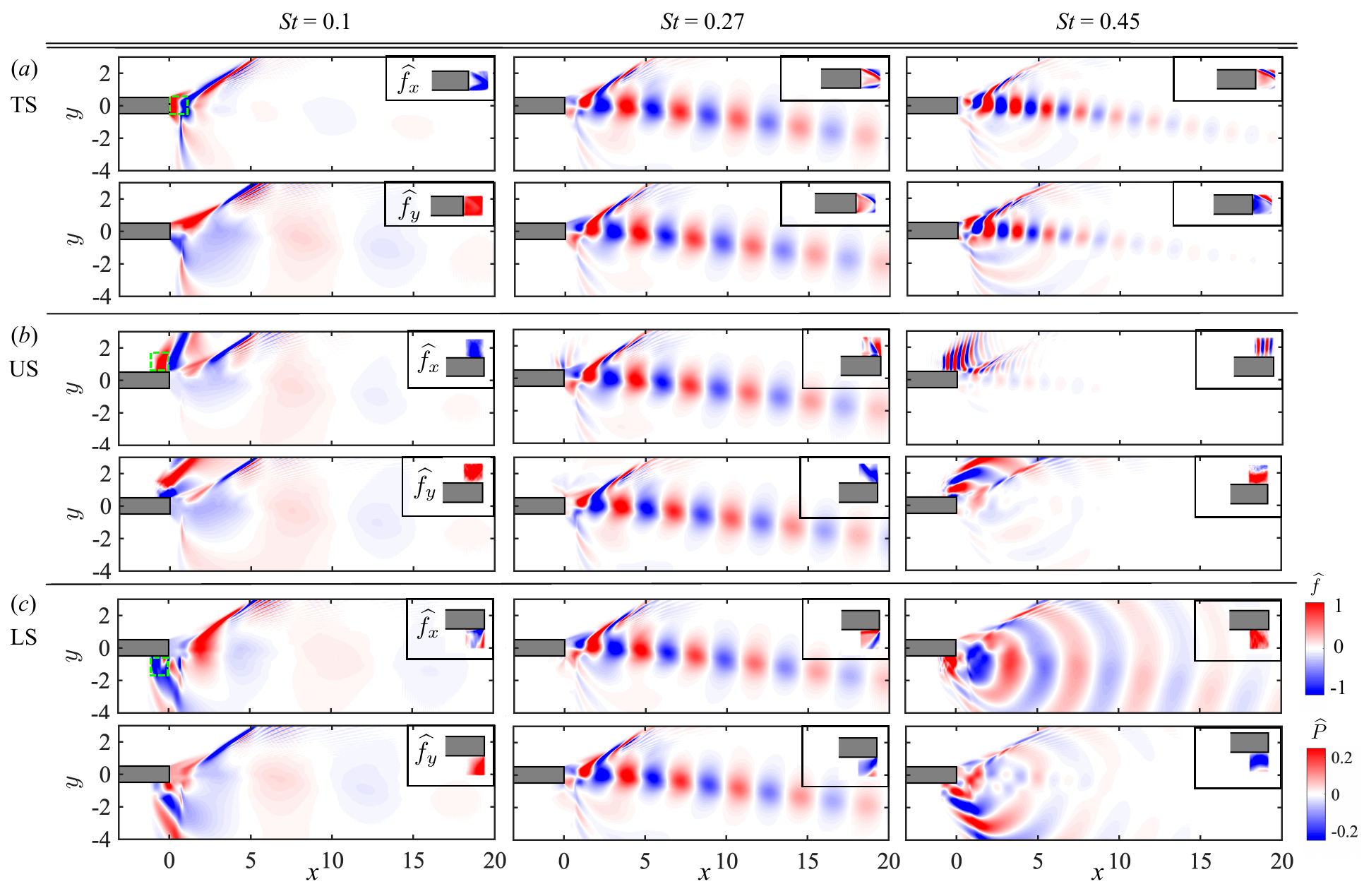}
    \caption{The spatial structures of the optimal input-output modes at three different frequencies: $St = 0.1$, 0.27, and 0.45. The top and bottom columns show the output pressure perturbation $\hat{P}$ for the $x-$direction $\hat{f}_x$ and $y-$direction $\hat{f}_y$ input, respectively, in each subplot. (a) Splitter plate trailing surface (TS), (b) upper surface (US), and (c) lower surface (LS) input locations. The green dashed lines denote the forcing location.}
    \label{fig:io_mode_velocity}
\end{figure}

Figure \ref{fig:io_mode_velocity} presents the spatial structures of optimal input-output mode pairs at three frequencies ($St$ = 0.1, 0.27, and 0.45) for different input locations. At the TS (Figure \ref{fig:io_mode_velocity}(a)), both $\hat{f_x}$ and $\hat{f_y}$ inputs exhibit similar responses in the pressure perturbation $\hat{P}$ across frequencies. The dominant response occurs in the shear-layer region, with additional responses observed along the lower side of the shear layer as propagating waves and along the oblique shock. The flow pattern is consistent with the phenomenon of stacked compression waves induced by the vortical structures convecting downstream captured in the LES.

Figure~\ref{fig:io_mode_velocity}(b) and ~\ref{fig:io_mode_velocity}(c) show the pressure output when the input location is constrained to the US and LS, respectively. At the dominant frequency $St = 0.27$, the pressure response is situated in the shear-layer region for both $\hat{f_x}$ and $\hat{f_y}$ inputs, with very weak waves observed on the upper side of the SPTE for both the input location. The shear-layer instability remains the most responsive mechanism. However, for frequencies deviated from the resonant frequency (i.e., lower frequency input at $St = 0.1$ and higher frequency at $St = 0.45$), response modes of pressure perturbation $\hat{P}$ concentrate near the forcing location. In contrast, the response in the shear layer is relatively weak. For both US and LS inputs, strong pressure waves emanating from the input location travel downstream, and their propagating direction aligns with the input forcing direction at $St=0.1,$ and 0.45. Interestingly, $St=0.45$ input in $\hat{f_x}$ and $\hat{f_x}$ at the US and LS mainly influence the pressure wave and do not impact the wavepackets in the shear layer. Low amplitude wavepackets are observed in the shear layer at the low frequency $St=0.1$ input.    

In summary, the shear layer exhibits a strong response to perturbation inputs applied at the splitter plate trailing surface across all frequencies. Input at the upper and lower surface locations introduces strong pressure waves emanating from the input location when the forcing frequency deviates from the resonant frequency of $St=0.27$. The pressure perturbation output is only consistently observed in the shear layer at the dominant frequency of $St=0.27$. This behavior is attributed to the convective nature of KH instabilities, indicating that regardless of spatial location and variable inputs, the shear layer is particularly responsive at $St = 0.27$. The analysis here also provides insights into the choice of input configuration to excite particular flow regions precisely.  

\subsubsection {Pressure as an input} \label{sec:io_pressure}

\begin{figure}
     \centering
         \includegraphics[width=0.75\textwidth]{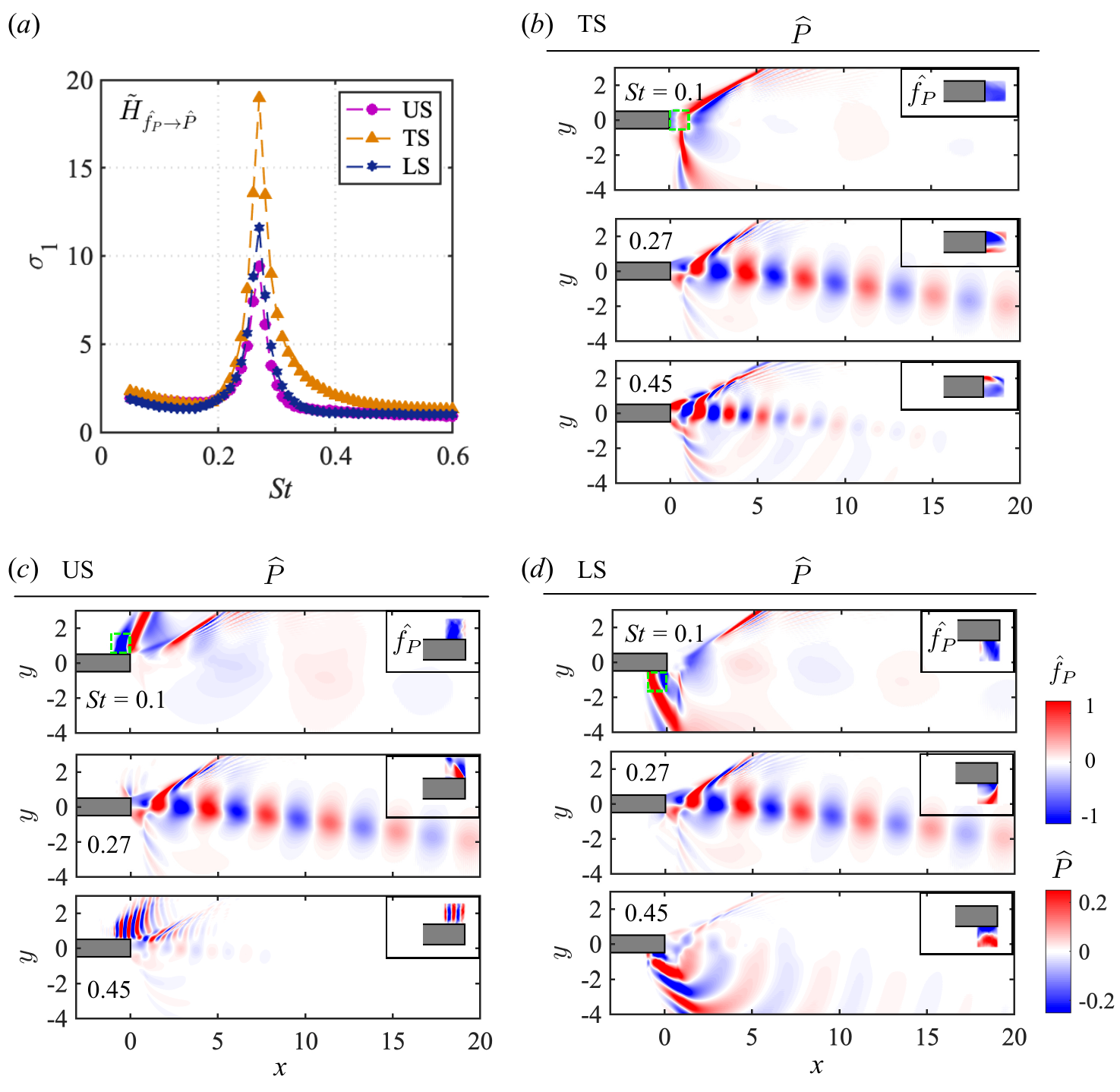}
    \caption{(a) $L2$-norm-based gain with pressure as input and output. Forcing is constrained at the upper surface (US), trailing surface (TS), and lower surface (LS) of the splitter plate. (b-d) Modal structures of the output and input pressure perturbation for three representative frequencies: $St = 0.1, 0.27$ and 0.45; (b) TS, (c) US, (d) LS spatial inputs.}
    \label{fig:io_pressure}
\end{figure}

In addition to the momentum-based forcing, we also examine pressure-based forcing, again using the pressure variable to characterize the output. Figure~\ref{fig:io_pressure}(a) illustrates that pressure-based forcing exhibits higher gains at the dominant frequency $St = 0.27$ across all locations considered. Specifically, the TS emerges as the optimal location for inducing the largest amplification compared to the US and LS. The energy amplification for the pressure-based input is relatively lower than the momentum-based forcing. These results indicate that momentum-based actuators may be more efficient in an active flow control design to affect the original flow behaviors. However, the results can be biased based on the norm used in the current study. The nonlinear simulations will be helpful in shedding light in this direction.

As shown in figure~\ref{fig:io_pressure}, the output pressure modal structure resembles the features observed in the cases when the forcing is applied in $x-$direction $\hat{f_x}$ for the respective input location and frequency. All the observations from the previous discussion on momentum-based input have also been captured for the pressure input. For example, the shear layer wavepackets and the pressure waves traveling far-field are responsive for all the frequencies while input is given at TS (see figure~\ref{fig:io_pressure}(b)). Due to the convective nature of the KH instabilities, the optimal pressure response is located in the shear layer at dominant frequency $St = 0.27$, regardless of the location of an input. A major distinct feature is noticed for the LS input at $St=0.45$ with pressure input compared to $\hat{f_x}$ input, in which the pressure waves traveling to the far field are located mainly in the bypass stream for input $\hat{f_P}$ (figure~\ref{fig:io_pressure}(d)), whereas those waves are in both the main and bypass stream for the input $\hat{f_x}$ (figure~\ref{fig:io_mode_velocity}(c)).        

\subsubsection {Isolating the far-field traveling waves} \label{sec:phase_speed}

The pressure output modes obtained for the different inputs are observed in wavepackets in the shear layer region or far-field traveling instability waves. These waves are linked to the acoustic far-field through a Mach-wave-like mechanism in supersonic flows (\citet{tam1995}). A small disturbance input placed upstream can trigger a supersonic instability wave downstream that contributes to noise generation. This instability process is well captured by parabolized stability equations (PSE) (\citet{rodriguez2013}) and input-output analysis (\citet{jeun2016}) for a jet flow.

In the present study, to compute the speed of wavepackets and instability waves, we consider a wave in the radial direction represented by $e^{-i k_r r}$ (\citet{karban2023}). Here, $k_r$ denotes the radial wavenumber, and the phase speed is given by $c_r = \omega/k_r$, where $\omega$ ($ = 2 \pi St U_\text{ref} / L $) is an angular frequency. The normalized projection between the first output pressure modes and this radial wave is determined by,
\begin{equation}
    \mathscr{B}_P =  \frac {\langle  \hat{P} (r, \theta, St) , ~e^{-i k_r r}  \rangle  }{\| \hat{P} \|_2 ~ \| e^{-i k_r r}\|_2} 
    \label{eq:proj_phase}
\end{equation}
The phase speed is varied $c_r/c_\infty = [0.1, 3]$, where $c_\infty$ is the ambient speed of sound. The center of the radial coordinate is approximately at $(x, y) = (0.75, 0)$.  

\begin{figure}
     \centering
         \includegraphics[width=0.6\textwidth]{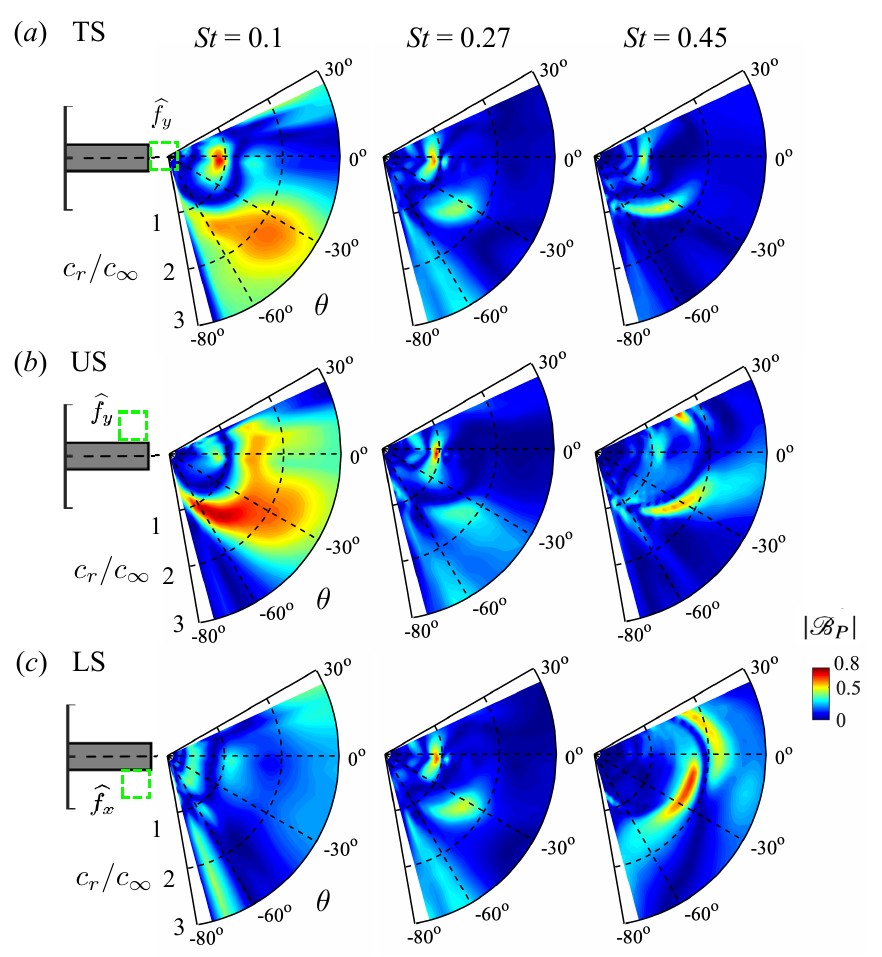}
    \caption{The absolute value of projection between the optimal pressure output mode and radial waves with varied phase speed $c_r/c_\infty = [0.1, 3]$ for three frequencies $St =$ 0.1, 0.27, and 0.45. (a) $\hat{f}_y$ input at the trailing surface (TS), (b) $\hat{f}_y$ input at the upper surface (US), and (c) $\hat{f}_x$ input at the lower surface (LS). A green box indicates an input location.}
    \label{fig:phase_speed}
\end{figure}

Figure~\ref{fig:phase_speed} illustrates the absolute value of the projection $\mathscr{B}_P$ for representative cases. At $St=0.27$ across all input locations and variables, the highest projection occurs at a phase speed of approximately $c_r/c_\infty \sim 0.8$ for shear layer wavepackets, characteristic of Kelvin-Helmholtz instability modes (\citet{schmidt2018}). Meanwhile, pressure instability waves in the bypass stream exhibit phase speeds greater than $c_\infty$, specifically around $c_r/c_\infty \sim 1.6$, indicative of Mach-wave-like mechanisms (\citet{ffowcs1963, crighton1975}). At a higher frequency of $St=0.45$, prominent pressure waves are observed for $\hat{f}_y$ input at US and $\hat{f}x$ input at LS (see figure~\ref{fig:io_mode_velocity}(b) and (c), respectively). These waves propagate at supersonic speeds in both the main and bypass streams, with phase speeds exceeding $c_r/c_\infty > 1$. Notably, for $\hat{f}x$ input at LS, the phase speed of pressure waves nearly exceeds $c_r/c_\infty > 2$, indicating potential significant far-field noise generation. At the low-frequency scenario of $St = 0.1$, the projection spans a broader range within $1.5 < c_r/c_\infty < 2.5$ in the bypass stream due to wider lobe structures associated with lower frequencies. Overall, the phase speed of pressure instability waves, which strongly influences far-field noise, is notably sensitive to both the input variable and its location.

\subsubsection {Effect of upstream input location} \label{sec:io_upstream}

\begin{figure}
     \centering
         \includegraphics[width=0.75\textwidth]{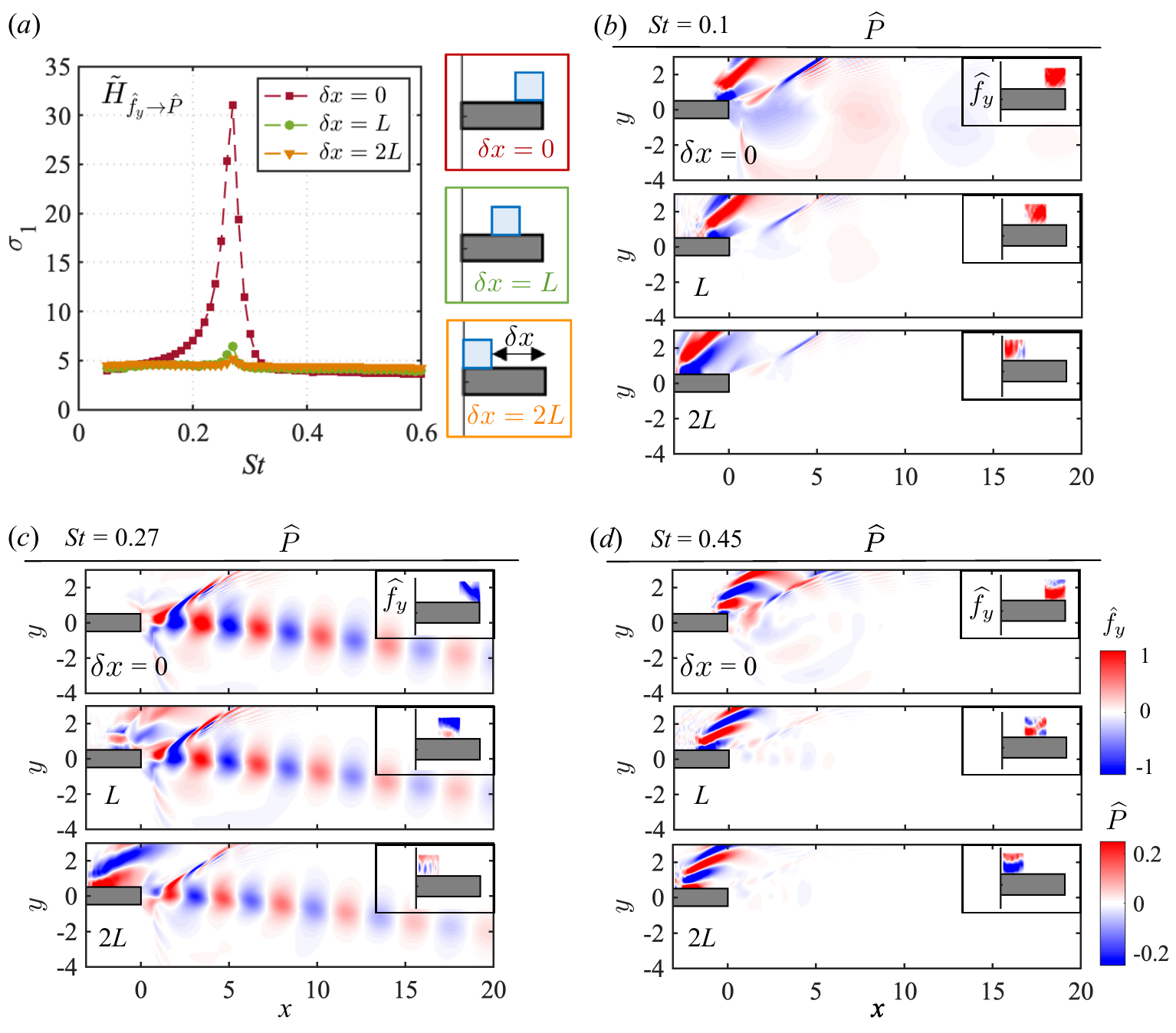}
    \caption{(a) $L2$-norm-based gain with $y$-direction forcing $\hat{f_y}$ as an input variable and pressure as output variable. The blue box indicates the location of an input. Modal structures of the output pressure and input $\hat{f_y}$ perturbations for three upstream locations at a frequency of (b) $St = 0.1$, (c) $St = 0.27$, and (d) $St = 0.45$.}
    \label{fig:upstream_v}
\end{figure}

In this section, we evaluate the effect of an input location on the pressure output response by moving the spatial location of the input upstream of the SPTE as shown in figure~\ref{fig:upstream_v}(a). Input is constrained at three locations on the upper surface with $\delta x = 0$, $L$, and $2L$, where $\delta x$ is the distance from SPTE.

The $y-$direction momentum forcing $\hat{f_y}$ is prescribed. As the input location moves upstream from the trailing edge, $\sigma_1$ is significantly reduced at $St=0.27$. However, $\sigma_1$ is almost identical for $St \lesssim  0.1$ and $St \gtrsim 0.35$. Figure~\ref{fig:upstream_v}(c) shows the shear layer wavepackets remain responsive at the upstream locations $\delta x = L$ and $2L$ mainly due to the dominating KH instability at $St = 0.27$. Additionally, the upstream input shows the strong pressure waves produced from the input. For $St = 0.1$ and $0.45$, the shear layer response becomes less dominant compared to the pressure wave response produced from the local input $\hat{f_y}$ as it is moved upstream, away from SPTE (figure~\ref{fig:upstream_v}(b) and (d)). 

\subsection {2-D nonlinear simulation with perturbations} \label{sec:2D_simulaiton}

Simulations are conducted for the baseline and perturbed supersonic shear layer flow using a compressible Navier-Stokes flow solver \textit{CharLES} (\citet{bres2017}). An unstructured grid is employed with a second-order finite-volume method and third-order Runge--Kutta time integration. The incoming flow conditions for the main and bypass streams are detailed in table~\ref{tab:1}. To mitigate reflected waves, a sponge layer is applied at the outlet and bottom boundaries, while the top boundary is treated with an inviscid wall condition to replicate the original 3-D LES dataset. The baseline, represented by a 2-D simulation without perturbations, is compared with the center plane of the 3-D LES data in appendix~\ref{sec:grid_study}, demonstrating a good agreement in the mean flow profile. Following a grid-independence study presented in appendix~\ref{sec:grid_study}, a grid comprising 0.76 million control volumes is selected for further investigation.

We have selected two cases of particular interest based on the linearized input-output analysis in the previous section for determining perturbation parameters. Notably, the upper surface forcing in the $y-$direction at frequencies $St=0.27$ and $0.45$ exhibit distinct features. At the dominant frequency $St=0.27$, the pressure wavepackets in the shear layer display a higher response, whereas the input at $St=0.45$ yields a dominant response in the pressure wave emanating from the US input location and propagating far-field as traveling waves (refer to figure~\ref{fig:io_mode_velocity}(b)). Therefore, we introduce perturbations at the upper surface of the splitter plate within a square box with $\delta x = 0$. A harmonic forcing is applied in the $y-$momentum equation, featuring a flux amplitude of approximately $12\%$ at two different frequencies, $St_a=0.2545$ and 0.45, where $St_a$ represents the frequency of a forcing.
 
\begin{figure}
     \centering
        \includegraphics[width=0.8\textwidth]{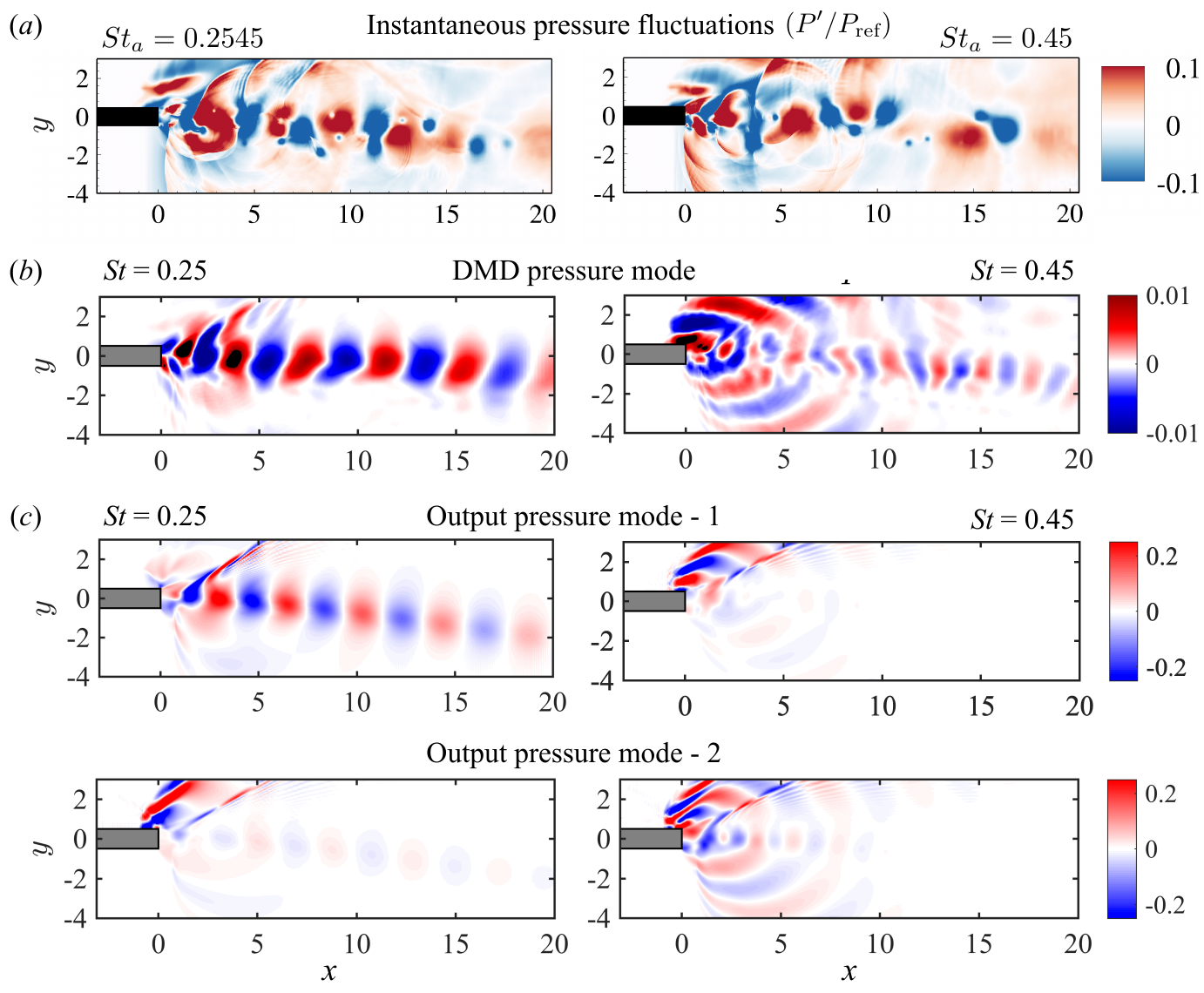}
    \caption{ 2-D simulations with a perturbation introduced in $y$-momentum equation at the upper surface of the splitter plate. (a) Instantaneous pressure fluctuation ($P'/P_{\text{ref}}$), (b) pressure DMD mode from the nonlinear simulation dataset, (c) The output pressure mode-1 and mode-2 predicted from linearized input-output analysis with the $\hat{f}_y$ forcing at the upper surface of the splitter plate.}
    \label{fig:2D_DMD}
\end{figure}

The instantaneous pressure fluctuations ($P'$) depict the strong pressure wave introduced by the forcing location for both cases, as shown in figure~\ref{fig:2D_DMD}(a). Since the instantaneous field contains coherent structures at multiple frequencies, identifying those associated with specific frequencies can be challenging. To address this, we employ the dynamic mode decomposition (DMD) method to extract the dynamically important mode. The DMD segregates time-resolved snapshots into modes, each associated with a single frequency (\citet{schmid2010, taira2017}). This is an appropriate choice for our analysis as it computes the mode related to a single frequency without a need for a long-time dataset, which is necessary for SPOD. The pressure flowfield is sampled with a time interval of $\triangle t U_{\text{ref}} / L \approx 0.19$, and $600$ snapshots are used to perform the DMD for both perturbed flow simulations. 

In figure~\ref{fig:2D_DMD}(b), the DMD pressure modes are presented, and figure~\ref{fig:2D_DMD}(c) depicts the corresponding output pressure modes obtained from input-output analysis. Significant similarities are observed between the DMD modes and output modes. The DMD pressure mode reveals higher fluctuations in the shear-layer wavepackets at $St=0.25$, accompanied by a low-amplitude pressure wave introduced from the forcing location. These features closely resemble those of the leading output mode-1. For the $St_a = 0.45$ case, the DMD mode exhibits very strong pressure waves originating from the forcing location, along with relatively low-amplitude pressure waves in both the main and bypass streams. These features align well with the leading output pressure mode-1 at $St=0.45$. Additionally, the wavepackets in the shear layer near SPTE, as evident in the DMD mode, are present in the output mode-2 with relatively low amplitude as compared to those pressure waves originating from the forcing location. 

Since DMD analysis does not provide a rank to its mode at each frequency, the dynamics captured by DMD can be observed as a superposition of the multiple output modes (\citet{towne2018}). We can further relate it to the amplification obtained from the linear analysis. Near the dominant frequency of $St = 0.25$, the leading energy amplification $\sigma_1^2$ is significantly higher than $\sigma_2^2$, i.e., $\sigma_1^2/\sigma_2^2 \approx \mathcal{O}(10) $, indicating the rank-1 behavior. Hence, the leading amplification mechanism -- wavepacket response in the shear-layer -- as predicted by the output pressure mode-1 is sufficient to capture the shear-layer dynamics as reflected by DMD mode. At $St=0.45$, the $\sigma_1^2$ and $\sigma_2^2$ are very comparable ($\sigma_1^2/\sigma_2^2 \approx \mathcal{O}(1) $), indicating that the suboptimal modes are equally important. In consequences, the features observed in the output mode-1 and mode-2 are superimposed in the DMD mode, where the pressure waves originating from the forcing mechanism are still the leading amplification mechanism. The downstream ($x \geq 10$) wavepackets, which may result from the nonlinear interactions, observed in DMD are not captured in all output modes. The absence of downstream shear-layer wavepackets identified in the DMD mode at $St = 0.45$ is attributed to nonlinear turbulence interactions beyond $x \geq 10$, which are not adequately captured by linear analysis methods. Despite this, it is noteworthy that predictions from the input-output analysis correlate effectively with nonlinear simulations within the region $x \le 10$.

\section{Conclusions} \label{sec:conclusion}

We use a combination of linear and data-driven techniques together with scale-resolving simulations to investigate the effect of perturbations on the development of a globally significant shear layer in a multi-stream rectangular nozzle flow. Previous studies have indicated that the mixing of the main ($M_\infty = 1.23$) and bypass ($M_\infty = 1.0$) streams near the trailing surface of the splitter plate cause high-frequency loading on the airframe and generate a prominent farfield noise signature. The baseline flow exhibits large-scale vortical structures that roll up and convect downstream along the shear layer due to Kelvin--Helmholtz (KH) instability. Spectral analysis reveals the dominant frequency of these structures at $St = 0.27$; at this frequency, spectral proper orthogonal decomposition (SPOD) of streamwise velocity displays KH instabilities along the shear layer, while modes of the pressure variable reveal wavepackets along the shear layer and pressure waves traveling in the main and bypass streams. In addition, bispectral mode decomposition (BMD) is utilized to reveal the dominant nonlinear energy exchanges. The BMD analysis illustrates the energy cascade from the KH instability to frequencies that exhibit significant energy in the SPOD apart from $St=0.27$.

The leading gain obtained from the classical resolvent, modified with a discounting parameter to account for the instability, highlights higher energy amplification at the dominant frequency, with an order of magnitude higher than the first sub-optimal gain. The forcing and response modes showcase the convective nature of instability at $St=0.27$. These observations corroborate the KH instability as the primary energy amplification mechanism in the supersonic shear layer flow. The interpretation of these dynamics is further explored by comparing SPOD and resolvent analyses.  A key observation is that the leading SPOD energy spectra and optimal resolvent energy amplification spectra follow similar trends. However, the resolvent gain spectrum misses peaks at the few energetic frequencies in SPOD; this is attributable to nonlinear interactions, as shown by BMD, in turbulent flows that are not captured in linear resolvent analysis. Moreover, the alignment of the leading SPOD and optimal resolvent response modes at the dominant frequency suggests that the coherent structures predicted by the linear analysis remain active in the highly turbulent flow.

The results are then used to guide componentwise input-output analysis by restricting inputs to feasible locations where the solution is sensitive to forcing, including the upper (US), lower (LS), and trailing (TS) surfaces. The effects of $x-$ and $y-$momentum, as well as pressure inputs (forcing), are explored on the output (response) pressure field. All combinations of state variables and spatial restrictions display higher amplification occurring at the dominant frequency and indicate the dominance of the KH instability as the primary amplification mechanism. The splitter plate trailing surface emerges as the most sensitive location for introducing perturbations. 

At the dominant frequency, the shear layer wavepackets and oblique shock exhibit high response regardless of the input location. The US and LS surface inputs introduce additional high-amplitude pressure waves originating from the input location in the main and bypass streams, respectively, while the shear layer shows a weaker response. The phase speed analysis reveals that the speed of the KH mode in the shear layer is 0.8 times the ambient speed of sound, consistent with past studies. Conversely, pressure waves traveling in the main and bypass streams convect at supersonic speed, highly influenced by the input location and variables. Moreover, shifting the US input further upstream of the SPTE results in a significant reduction in amplification at the dominant frequency. The pressure response shifts to high-amplitude pressure waves originating from the input location. At the same time, the wavepackets in the shear layer are weaker except at the dominant frequency due to the prominent effect of the KH instability.

Finally, demonstrative 2-D simulations are conducted with unsteady forcing applied to the upper surface of the splitter plate at two different frequencies. The pressure modes obtained through dynamic mode decomposition (DMD) closely match the pressure output modes predicted by the input-output analysis at the corresponding frequency. The optimal output pressure mode alone accurately predicts the prevailing flow features observed in the DMD pressure mode with the input at the dominant frequency; whereas, for the $St=0.45$ forcing case, both the optimal and second pressure modes are equally essential for capturing the overall dynamics, as shown by comparing these modes with the DMD pressure mode. In summary, we have demonstrated that linear input-output analysis effectively captures the dominant dynamics of highly nonlinear turbulent supersonic shear layer flow. The insights gleaned from the input-output analysis can be leveraged to design effective flow control strategies in supersonic shear-layer flow scenarios.

\section*{Acknowledgements}

We gratefully acknowledge support from the Air Force Office of Scientific Research (AFOSR) under award FA9550-23-1-0019 (Program Officer: Dr. Gregg Abate). The main LES studies were carried out using resources provided by the U.S. Department of Defense High-Performance Computing Modernization Program and the Ohio Supercomputer Center. We also acknowledge the computational resources provided by Syracuse University Research Computing. The authors would like to thank Dr. Mark N. Glauser for fruitful discussions.
~\\

\appendix

\section{Grid independence study}
\label{sec:grid_study}

Table~\ref{tab:grid} displays the first singular value of the resolvent operator for $Re=500$ and $St=0.27$ with discounted parameter $\alpha=0$. The percentage change in $\sigma_1$ is less than 2\% for all three grids, indicating that the flow mechanism associated with the leading gain has been well resolved. However, the forcing and response modal structures (not presented here) exhibit grid dependency in the shear layer $x > 10$ for the case with grid G2. To ensure grid resolution for capturing dominant modes, local grid refinement is employed in the shear-layer region in the G3 grid. The forcing and response modes calculated on this grid show no grid dependency, and we use the grid G3 for all the analyses in the present work.
\begin{table}
\begin{small}
\centering
\begin{tabular}{l l l l} 
\hline
Grid & $St$ & $\sigma_1$ & \% change \\ [0.5ex] 
\hline\hline
G1: 0.101 M & 0.27 & $2.2982 \times 10^3$ & -- \\ 
G2: 0.136 M & 0.27 & $2.2856 \times 10^3$ & -0.55\% \\
G3: 0.176 M & 0.27 & $2.2554 \times 10^3$ & -1.34\% \\
 \hline
\end{tabular}
\caption{The first singular value ($\sigma_1$) at the dominant frequency $St = 0.27$ for different grid resolutions.}
\label{tab:grid}
\end{small}
\end{table} 

\begin{figure}
     \centering
         \includegraphics[width=0.9\textwidth]{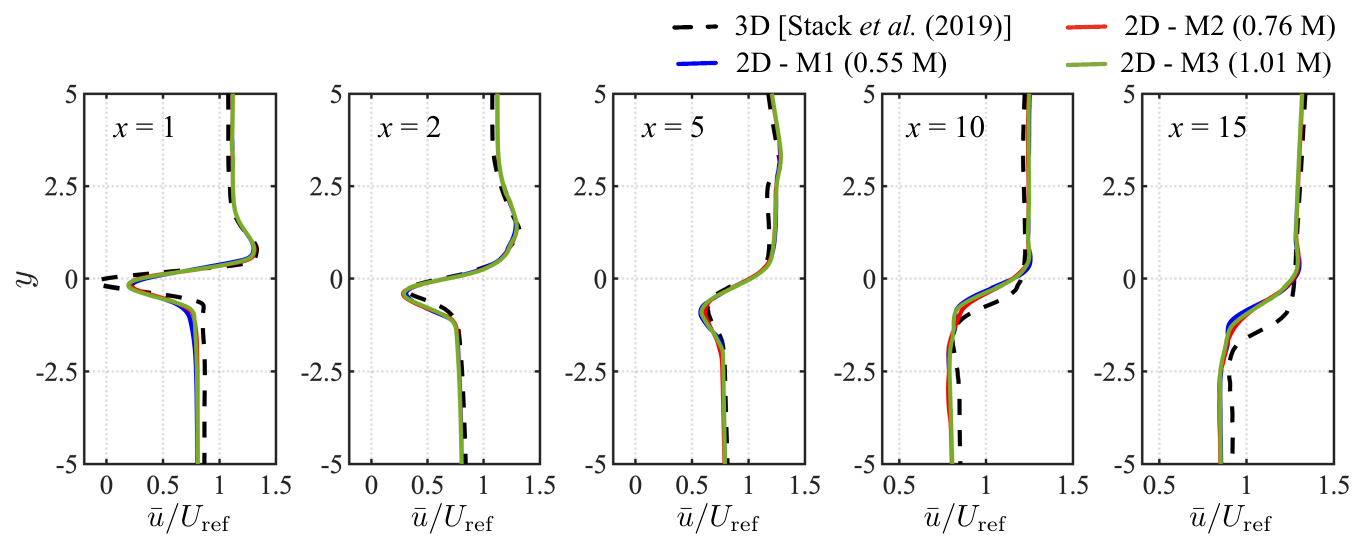}
    \caption{A comparison of a normalized time-averaged streamwise velocity $\overline{u}/U_{\text{ref}}$ at various streamwise locations ($x=1, 2, 5, 10$, and 15) between the center plane taken from 3-D and 2-D simulations with different grids. "M" represents a million.}
    \label{fig:3D_mean_grid}
\end{figure}

The 2-D simulations are also conducted using three grids, namely M1 (0.55 million cells), M2 (0.76 million cells), and M3 (1.01 million cells). For all three grids, the $y+$ value is maintained at approximately 1. The simulations are performed for a convective time $tU_{\text{ref}}/L \approx 267$ after the initial transition phase. Figure~\ref{fig:3D_mean_grid} compares the normalized time-averaged streamwise velocity $\overline{u}/U_{\text{ref}}$ at various streamwise locations ($x = 1,2,5,10,$ and 15) for all three grids. A minor difference is observed in $\overline{u}$ between grid M1 and M2 in the shear-layer region, while $\overline{u}$ overlaps for M2 and M3. The flow structures are well resolved for grid M2 in the shear-layer region, and no significant changes are observed for grid M3. Consequently, we opt to utilize grid M2 with 0.76 million cells for the 2-D nonlinear simulations.

We also compared the results obtained from the 2-D simulations and the center plane data from the 3-D simulations performed by \citet{stack2019splitter} in figure~\ref{fig:3D_mean_grid}. The time-averaged streamwise velocity shows a good agreement between 3-D and 2-D cases overall. In the vicinity of the SPTE ($x=1$), the 2-D time-averaged streamwise velocity slightly underpredicts the sharp change in the shear-layer region, while a reasonable good match is demonstrated between the 2-D and 3-D cases downstream of the SPTE ($x \le 10$). However, a deviation is observed in the $\overline{u}$-velocity far away from the SPTE at $x=15$. The breakdown of the 2-D roller structures produced at the SPTE as they move downstream leads to a spanwise variation in the flowfield, potentially causing the observed deviation in the 2-D and 3-D simulations' mean flowfields far from the SPTE. Despite this, since we achieved a good agreement in the vicinity of the splitter plate between the 2-D and 3-D simulations, it is reasonable to use the 2-D simulations to validate the predictions obtained from the input-output analysis. It is worth noting that the dominant frequency ($St = 0.273$) reported in the 3-D simulation slightly changes to $St=0.2545$ in the 2-D simulation, but this change does not lead to a significant change in the dominant flow mechanism.

\section{Randomized singular value decomposition} \label{sec:rsvd_method}

The algorithm used in the present study to perform randomized singular value decomposition is presented in this section. Instead of performing SVD on $\boldsymbol{H}(\boldsymbol {Q}; \omega,  \beta)$, a low-rank representation of the resolvent operator can be derived by finding an appropriate low-dimensional basis to project the large resolvent operator on a suitable subspace. A \textit{sketch} $\boldsymbol{S}$ of the operator $\boldsymbol{H}(\boldsymbol {Q}; \omega,  \beta)$ is obtained by using a tall and skinny test matrix $\Upsilon$ as 
\begin{equation}
\boldsymbol{S} = \boldsymbol{H}(\boldsymbol {Q}; \omega,  \beta) \boldsymbol{\Upsilon},
\end{equation}
where $\boldsymbol{\Upsilon} \in \mathbb{R}^{m \times k}$, $m$ is the size of the resolvent operator, $k$ is total number of desired singular values and $k \ll m$.
A sketch matrix $\boldsymbol{S}$ contains the dominant information of $\boldsymbol{H}(\boldsymbol {Q}; \omega,  \beta)$ (\citet{halko2011, woolfe2008, tropp2017}). The test matrix $\boldsymbol{\Upsilon}$ can be constructed using random values with normal Gaussian distribution (\citet{martinsson2011}) or be weighted by some physics-informed input matrix (\citet{ribeiro2020}). As the sketch holds the dominant features of $\boldsymbol{H}(\boldsymbol {Q}; \omega,  \beta)$, we can form an orthonormal basis with $\boldsymbol{O} \in \mathbb{C}^{k \times m}$ by using QR decomposition of $\boldsymbol{S}$. Then, the projection of $\boldsymbol{H}(\boldsymbol {Q}; \omega,  \beta)$ onto $\boldsymbol{O}$ is obtained to derive its low-rank approximation such that $\boldsymbol{H}(\boldsymbol {Q}; \omega,  \beta) \approx \boldsymbol{O M}$, where $\boldsymbol{M} = \boldsymbol{O}^* \boldsymbol{H}(\boldsymbol {Q}; \omega,  \beta) \in \mathbb{C}^{k \times m} $. Next, we perform the SVD upon this reduced-size matrix $\boldsymbol{M}$ as
\begin{equation}
\boldsymbol{M} = \boldsymbol{\Tilde{U} \Sigma V}^*,
\end{equation}
then the low-rank approximation of the resolvent operator is given by, 
\begin{equation}
\boldsymbol{H}(\boldsymbol {Q}; \omega,  \beta) \approx \boldsymbol{O \Tilde{U} \Sigma V}^*,
\end{equation}
where the left singular vector is approximated as $\boldsymbol{U = O \Tilde{U}}$ (\citet{halko2011}), and the right singular vector is given by $\boldsymbol{V}$. 

\begin{figure}
     \centering
         \includegraphics[width=0.95\textwidth]{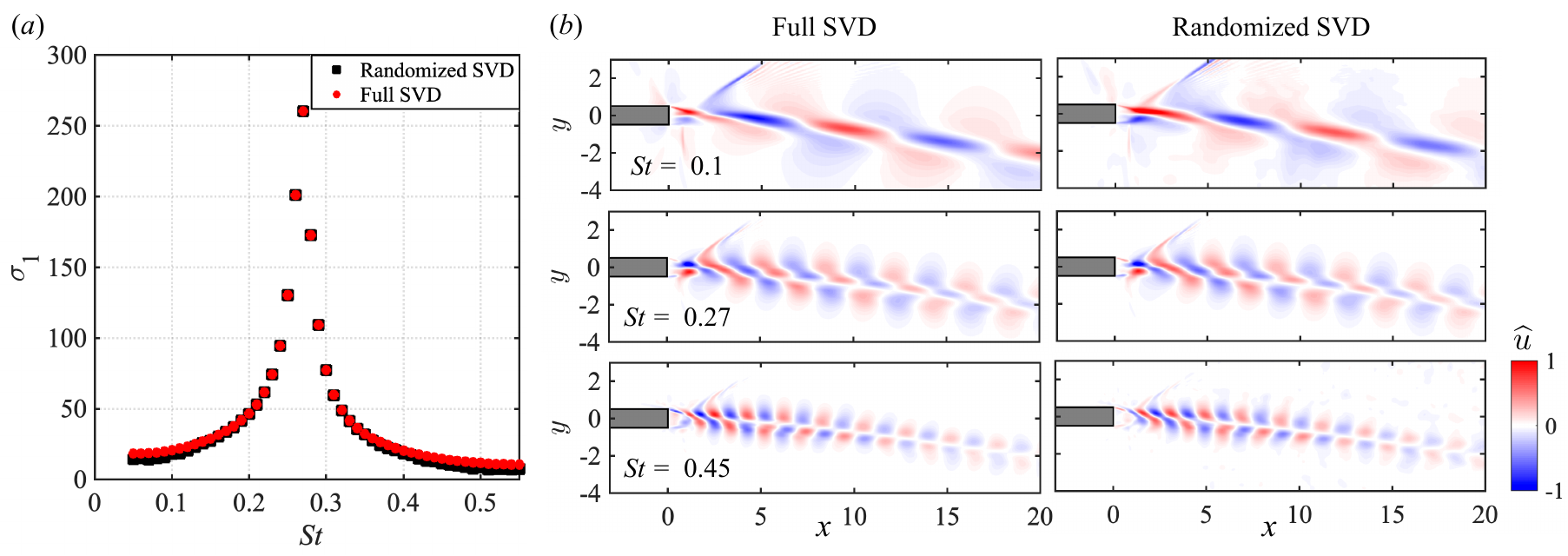}
    \caption{(a) The first singular value ($\sigma_1$) of randomized SVD and full SVD with $\alpha L/U_{\text{ref}} = 0.1$. (b) The streamwise velocity component ($\hat{u}$) of the leading response modes from full and randomized SVD for $St = 0.1, 0.27$, and 0.45.}
    \label{fig:svd_vs_rsvd}
\end{figure}

In the present study, a test matrix with $k=11$ (where $k$ is a total number of desired singular values) is used to compute the SVD of $\boldsymbol{H}(\boldsymbol {Q}; \omega, \beta=0)$, which results in a significant reduction in the dimension of the matrix. The comparison of the results with $\alpha L / U_{\text{ref}} = 0.1$ obtained from randomized SVD and full SVD (Arnoldi method, MATLAB \textit{svds}) is shown in figure~\ref{fig:svd_vs_rsvd}. We observe remarkable agreement for the leading singular value $\sigma_1$ and mode structures as shown in figure~\ref{fig:svd_vs_rsvd}. This justifies using randomized SVD to enable the processing of relatively large input-output operators in the present work.

\section{Reynolds number effects}
 
\begin{figure}
     \centering
         \includegraphics[width=1\textwidth]{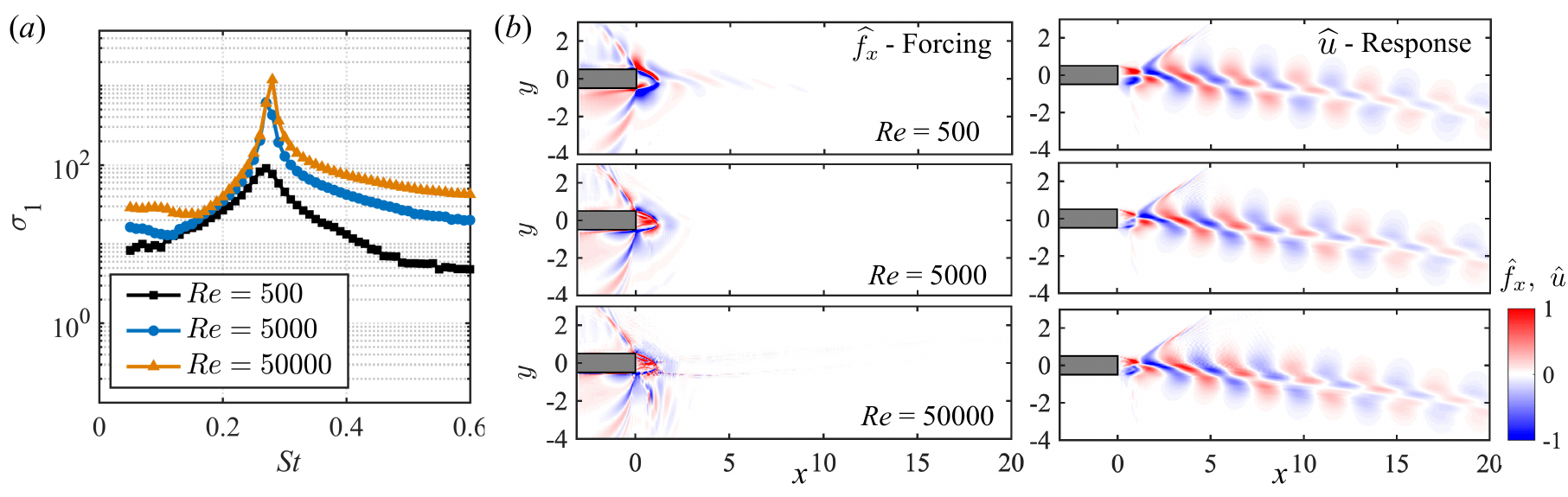}
    \caption{(a) The leading gain $\sigma_1$ distribution for the representative Reynolds numbers, and (b) the $x-$direction forcing and response mode pair for $St=0.27$ for different Reynolds numbers.}
    \label{fig:Re_study}
\end{figure}

In this section, we investigate the influence of Reynolds number on the leading forcing and response modal structures and their amplification ratio. Previous studies have explored the impact of Reynolds numbers on the linearized Navier-Stokes operator (\citet{schmidt2017}) and the resolvent operator (\citet{schmidt2018, doshi2022}). Figure \ref{fig:Re_study}(a) depicts the leading resolvent gain for three different Reynolds numbers, spanning an order of magnitude. Prior to conducting the resolvent analysis, a stability analysis is performed for all Reynolds numbers to determine the appropriate discounted parameter. It is observed that the growth rate of the unstable eigenvalue at $St = 0.27$ increases with Reynolds number. The discounted parameter is set as $\alpha L / U_\text{ref} = 0.15$ to compute the resolvent gain for all Reynolds numbers, ensuring that the most unstable eigenvalue resides below it. While the gain for all frequencies increases with the Reynolds number, the overall gain distribution trend remains similar. Moreover, the amplification mechanism does not change with a Reynolds number (Figure~\ref{fig:Re_study}); we consider the Reynolds number as a free parameter (\citet{schmidt2018}). 

Figure~\ref{fig:Re_study}(b) shows the $x$-direction forcing and response pair at $St=0.27$ for the representative Reynolds number. The response modal structures resemble the KH instability for all three Reynolds numbers and exhibit no qualitative discrepancies. These observations align with the findings of a previous study by \citet{pickering2021}, which concluded that inviscid instabilities are independent of the eddy-viscosity model at a sufficiently high Reynolds number.  
The forcing mode comprises noise-type structures in the shear layer for high $Re=50000$. It is noteworthy that further grid refinement does not result in cleaner forcing modes. The grid requirement to compute clean forcing modes at such high Reynolds numbers might be even higher than the LES grid. However, the forcing modal structures do not exhibit significant differences for all Reynolds numbers. The present study focuses on inviscid-type KH instability, which dominates near- and far-field fluctuations as discussed in \S~\ref{sec:baseflow}. Therefore, to reduce computational costs without altering the physics of the flow, we choose $Re=500$ for our linear analysis.

\section{Discounted parameter effects} \label{sec:dis_effect}

\begin{figure}[hbpt]
     \centering
         \includegraphics[width=0.5\textwidth]{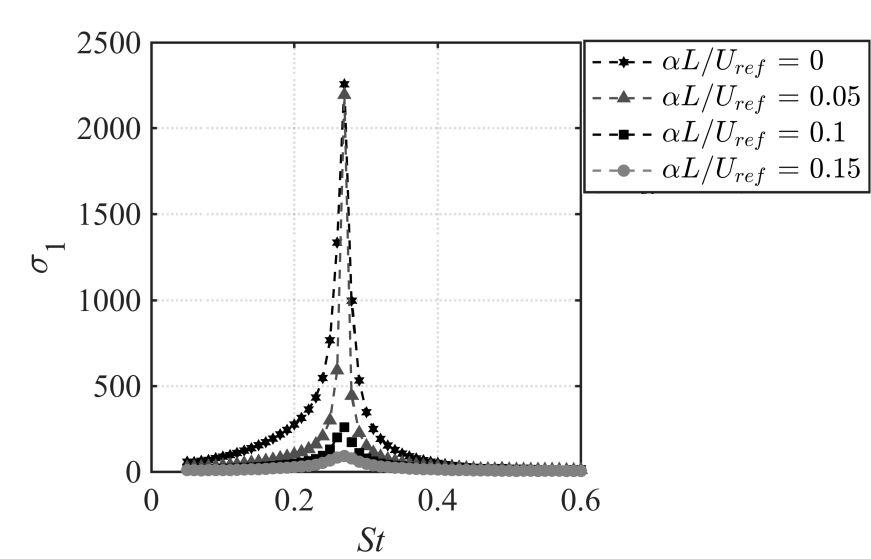}
    \caption{The leading gain $\sigma_1$ distribution for the representative discounted parameter ($\alpha L/U_{\text{ref}}$) from 0 to 0.15.}
    \label{fig:discounted_study}
\end{figure}

The appendix reports the effect of a discounted parameter ($\alpha L / U_{\text{ref}}$) on the leading gain $\sigma_1$ distribution. The overall trend of $\sigma_1$ remains consistent, with a peak observed at $St = 0.27$ for all $\alpha L / U_{\text{ref}}$ values considered as shown figure~\ref{fig:discounted_study}. When the discounted parameter is around the growth rate of $\omega_i L /U_{\text{ref}} = 0.0365$ at $St=0.27$ (i.e., $\alpha L /U_{\text{ref}} = 0$ or 0.05), the leading gain exhibits a large value due to a resonance between the input frequency and the unstable instability (figure~\ref{fig:stability}). For the higher values of $\alpha L /U_{\text{ref}} = 0.1$ and 0.15, the $\sigma_1$ reduces at all frequencies but more significantly at $St=0.27$. This analysis indicates that the identification of peak frequency $St = 0.27$ is independent of the select discounted parameter range, and other peaks are not observed by choosing other values of $\alpha L /U_{\text{ref}}$.

\bibliographystyle{plainnat}
\bibliography{references}

\end{document}